\documentclass{elsart}

\usepackage{graphicx}

\usepackage{amssymb}
\usepackage{amsmath}

\newtheorem{proposition}{Proposition}[section]

\begin{document}

\begin{frontmatter}



\title{Transport Induced by Mean-Eddy
 Interaction: \\
 I. Theory, and Relation to Lagrangian Lobe Dynamics}
\author[UMD]{Kayo Ide}\ead{ide@umd.edu}
\ead[url]{http://www.atmos.umd.edu/~ide}
 and
\author[UOB]{Stephen Wiggins} \ead{S.Wiggins@bris.ac.uk}
\ead[url]{http://www.maths.bris.ac.uk/people/faculty/maxsw/}
\address[UMD]{Department of Atmospheric and Oceanic Science, \\
Center for Scientific Computation and Mathematical Modeling, \\
Institute for Physical Science and Technology, \\ \&
Earth System Science Interdisciplinary Center, \\
University of Maryland, College Park, USA}
\address[UOB]{School of Mathematics, University of Bristol, Bristol BS8 1TW, UK}

\begin{abstract}

In this paper we develop a method for the estimation of 
{\bf T}ransport {\bf I}nduced by the {\bf M}ean-{\bf E}ddy 
interaction (TIME) in two-dimensional unsteady flows.
The method is 
built on the dynamical systems approach
and can be viewed as a hybrid combination of Lagrangian and Eulerian 
methods.
The (Eulerian) boundaries across which we
consider (Lagrangian) transport are kinematically defined by
appropriately chosen streamlines of the mean flow. By 
evaluating the impact of the mean-eddy interaction on 
transport, the TIME method can be used as a diagnostic tool for
transport processes that occur during a specified
time interval along a specified boundary segment.

We introduce two types of TIME functions:  one that quantifies the
accumulation of flow properties and another that measures the
displacement of the transport geometry. The spatial geometry of
 transport is described by the so-called pseudo-lobes, and temporal
 evolution of transport by their dynamics.
In the case where the TIME
functions are evaluated along a separatrix, the
pseudo-lobes have a relationship to the lobes of Lagrangian
transport theory. In fact, one of the TIME functions is identical
to the Melnikov function that is used to measure the distance, at leading order in a small parameter,
between the two invariant manifolds that define  the Lagrangian lobes. 
We contrast the similarities and differences
between the TIME and Lagrangian lobe dynamics in detail. An
application of the TIME method is carried out for inter-gyre transport in
the wind-driven oceanic circulation model and a comparison with
the Lagrangian transport theory is made.

\end{abstract}

\end{frontmatter}

\newpage
\renewcommand{\baselinestretch}{0.9}
\tableofcontents
\renewcommand{\baselinestretch}{1}
\newpage

\newcommand{\fginsrt}[1]{\marginpar{[Fig.\ref{#1}]}}
\newcommand{\tblinsrt}[1]{\marginpar{[Tab.\ref{#1}]}}


\section{Introduction and motivation}
\label{sec:intro}

Lagrangian transport methods are based on following the individual
trajectories obtained by solving the original differential equation
(ODE) for the particle location
starting from a set of initial conditions
${\bf x}_{0}$ at time $t_{0}$:

\begin{eqnarray}  \label{eq:dxpdt}
\frac{d}{dt} {\bf x} = {\bf u} ({\bf x},t)~,
\end{eqnarray}
\noindent 
where ${\bf u}$ is the velocity field. The
geometrical approach of  dynamical systems theory is particularly useful
when the flow field has Lagrangian coherent structures that separate the
flow into distinct regions. Then Lagrangian lobe dynamics describes the
transport process between these regions using stable and unstable
manifolds of hyperbolic trajectories as (moving) boundaries. The
Lagrangian methods have been applied successfully to a number of
unsteady geophysical flow problems; for a review, see
\cite{wiggins_arfm05,samwig}. If the flow is steady, i.e.,
${\bf u}({\bf x},t)=\overline{{\bf u}}({\bf x})$, the invariant
manifolds are stationary and no transport occurs between the regions.

On the contrary, Eulerian-based methods are mainly concerned with
the amount of transport across  stationary (Eulerian) boundaries
without computing  individual trajectories. An advantage of
Eulerian methods is that they tend to be much less elaborate than
Lagrangian methods in terms of computational implementation. 
The choice for the Eulerian boundaries is generally flexible, 
unlike the Lagrangian methods. 

From the dynamical systems point of view,
a parallel development of a method that computes
transport across the Eulerian boundary has yet to  take place. 
In this paper we begin the development of such a method.
The method  makes use of the interaction between the
reference (mean) state and the unsteady variability (eddy) as the
fundamental mechanism of transport. 
Hence we refer to it as the {\bf T}ransport {\bf I}nduced by the 
{\bf M}ean-{\bf E}ddy interaction (TIME). 
Using a streamline of the reference state as the boundary 
across which we consider transport,
TIME can be thought as a hybrid of  Lagrangian and Eulerian methods. 
Like the Eulerian method, the boundary  is stationary.
Like the Lagrangian method, the boundary is kinematically defined
and there is no TIME in the steady flow without 
the unsteady eddy component in the velocity. 
In certain situations we are able to
describe the geometrical relationship of TIME along the Eulerian
boundaries with Lagrangian lobe dynamics.

We require no assumption of incompressibility in our theoretical
framework. Therefore the ideas and techniques of the TIME method
can be applied to two-dimensional compressible flow or
three-dimensional volume-preserving flow which can be represented
as special classes of two-dimensional flows, such as the
shallow-water model. Remarks concerning incompressibility are
provided throughout the paper as special cases. Extensions to
three-dimensional flow are possible \cite{wiggins_ide_prep06}, but
there is more complexity in the geometry of the transport, and this is will be the topic of a future publication.

The outline of this paper is as follows. In Section~\ref{sec:bg}, we
provide a brief mathematical background and introduce the notion
of a kinematically-defined Eulerian boundary; readers who are
familiar with elementary dynamical systems theory may omit this
section without significant loss of continuity by referring back
to the notation and definitions as necessary. 
A brief glossary is also provided in Table~\ref{tbl:glssry}.  
The TIME method is defined in
Section~\ref{sec:fnc}, along with the two types of TIME functions. 
These functions, along with the notion of pseudo-lobes, are further 
explored  in Section~\ref{sec:chr}. 
An application of the TIME method is carried out in
Section~\ref{sec:application} for the inter-gyre transport in the
double-gyre ocean circulation model 
and a comparison with Lagrangian transport theory is presented.
Appendix~\ref{sec:P} provides details of perturbation theory,
and  Appendix~\ref{sec:L} compares the TIME method  with the
Lagrangian transport methods. 
\tblinsrt{tbl:glssry}

While in this paper we focus on introducing and developing the two
TIME functions that estimate the amount and the geometry of transport,
in the companion paper \cite{ide_wiggins_pd06b}, we expand the TIME
method further as a diagnostic tool for transport processes
by analyzing in detail the
influence of the mean-eddy interaction.

\section{Mathematical background}
\label{sec:bg}

In this section, we introduce the basic mathematical background
necessary to develop the TIME method.
The starting point is
first expressing the velocity field (\ref{eq:dxpdt}) in the following
form:

\begin{eqnarray} \label{eq:uunstdy}
   {\bf u} ({\bf x},t) = \overline{{\bf u}} ({\bf x}) +  {\bf u}'({\bf x},t)
\end{eqnarray}

\noindent where $\overline{\{\cdot\}}$ and $\{\cdot\}'$,
respectively, correspond to the steady reference state and the
unsteady fluctuation around the reference state. The choice of
reference state may not be unique. We choose the time-average
(mean) of the full time-dependent field as the reference state in
this study because the mean-eddy decomposition is natural when the
flow field is given by a data set; the TIME method itself does not
require the reference state to be the mean. Many of our results
will be perturbative in nature, with the (small) perturbation
parameter being the amplitude of the fluctuation that is
implicitly included in ${\bf u}'({\bf x},t)$ with respect to
$\overline{{\bf u}} ({\bf x})$. Appendix~\ref{sec:pert_just} gives results
on the length of time intervals on which perturbed trajectories
remain close to trajectories of the reference state.
These results will provide the validity of the
perturbative nature of our method since the TIME functions that we
derive will be of the form of integrals along perturbed
trajectories and approximations that, in principle, can be analytically
computed are of the form of integrals along trajectories of the
reference state. The regularity assumptions required on the velocity
field are minimal. Essentially, we need existence and uniqueness of
fluid particle trajectories, the ability to linearize about points in
space, compute Taylor expansions through second order with respect to
parameters, and for certain integrals of components of the velocity
field along trajectories of the reference velocity field to
exist. Assuming that the velocity field is twice continuously
differentiable with respect to the spatial coordinates, time, and any
parameters is adequate. No further assumptions on the nature of the time
dependence (e.g. time periodicity, quasiperiodicity, etc.) are required.

The use of perturbation theory in the development  of the TIME method is
made more transparent if we introduce an ''order parameter'',
$\varepsilon$,  associated with the fluctuation term as follows:

\begin{eqnarray}\label{eq:uhat}
{\bf u}'({\bf x},t)=\varepsilon \hat{{\bf u}}({\bf x},t)
\end{eqnarray}

\noindent where
$|{\bf u}'(\overline{{\bf  x}}^{C}(s),t)|
={\cal O}(\varepsilon|\overline{{\bf u}}(\overline{{\bf x}}^{C}(s))|)$.
The introduction of $\varepsilon$ in this way makes perturbation
arguments more transparent.
However, the TIME functions can be equally as well expressed in terms of
${\bf u}'({\bf x},t)$ or  $\varepsilon \hat{{\bf u}}({\bf x},t)$,
but in either case the approximation is to leading order in the size of
the fluctuation.

\subsection{Reference state and kinematically-defined Eulerian boundary}
\label{sec:bg_stdy}

We refer to a curve as Eulerian if it is stationary.
The TIME method uses an Eulerian curve
$C=\{\overline{{\bf x}}^{C} (s)\}$  that
is defined kinematically as a streamline of the reference flow.
It can be given as a solution of

\begin{eqnarray} \label{eq:s-def}
    \frac{d}{ds} \overline{{\bf x}}^{C} (s)
  &=&\overline{{\bf u}} (\overline{{\bf x}}^{C} (s))~
\end{eqnarray}

\noindent
with an initial condition ${\bf x}_{0}=\overline{{\bf
x}}^{C}(s_{0})$ at time $t_{0}$. For $C$ to be a physically
meaningful boundary, $\overline{{\bf x}}^{C}(s_{0})$ must be a
regular point of $\overline{{\bf u}}({\bf x}_{0})$, i.e.,
$|\overline{{\bf u}}({\bf x}_{0})|\neq 0$.

A trajectory with an initial condition $\overline{{\bf x}}^{C}(s_{0})$ at
time $t_{0}$ reaches $\overline{{\bf x}}^{C}(s_{0}-t_{0}+t)$ at time $t$
in the reference flow.
This trajectory is uniquely
identified  by a scalar, $s_{0}-t_{0}$, because
time shifts of a trajectory remain on the same trajectory in the reference
flow.
Throughout the paper, we interpret
the \emph{flight-time coordinate variable} $s$
strictly as a spatial coordinate variable along
$C$ while $t$ is a temporal variable.
Accordingly
$(\overline{{\bf x}}^{C}(s),t)=(\overline{{\bf x}}^{C}(s_{0}-t_{0}+t),t)$
and

\begin{eqnarray} \label{eq:xmn-sgm}
  (s,t)=(s_{0}-t_{0}+t,t)
\end{eqnarray}

\noindent
can be viewed as different parametrizations of the same
trajectory, which we call the \emph{reference trajectory}. The
Cartesian pair of coordinates $(s,t)$ will prove to be
particularly convenient for describing the TIME method. A glossary
is provided in Table~\ref{tbl:glssry} for the principal
definitions.

A hyperbolic stagnation point is a singular point.
In the reference flow, it is a special trajectory called the
\emph{distinguished hyperbolic trajectory} (DHT)
 \cite{ide_etal_npg02}
and we denote it by $\overline{{\bf x}}^{{\rm dht}}$,  i.e.,
$|\overline{{\bf u}}(\overline{{\bf x}}^{{\rm dht}})|=0$.
Although $\overline{{\bf x}}^{{\rm dht}}$ itself cannot be a physically
meaningful Eulerian boundary $C$,
the unstable and stable invariant manifolds that have a DHT
at the starting and end point, respectively, are special types of
$C$:

\begin{subequations} \label{eq:WUS}
\begin{eqnarray}
   \overline{W}^{U}&=&
   \{\overline{{\bf x}}^{U}(s)|\lim_{s\to-\infty}
    \overline{{\bf x}}^{U}(s)=\overline{{\bf x}}^{{\rm dht}}\};
    \label{eq:WUS-U} \\
   \overline{W}^{S}&=&
   \{\overline{{\bf x}}^{S}(s)|\lim_{s\to \infty}
   \overline{{\bf x}}^{S}(s)=\overline{{\bf x}}^{{\rm dht}}\},
    \label{eq:WUS-S}
\end{eqnarray}
\end{subequations}

\noindent
where $\overline{W}$ denotes  the reference manifold
with the superscripts $\{\cdot\}^{U}$  and $\{\cdot\}^{S}$ for
unstable and stable invariant manifolds, respectively. In a case
where $\overline{W}^{U}$ and $\overline{W}^{S}$  coincide, the
reference streamline is called the \emph{separatrix} or
\emph{heteroclinic connection} of the upstream DHT $\overline{{\bf
x}}^{{\rm dht}}_{-\infty}$ and the downstream DHT $\overline{{\bf
x}}^{{\rm dht}}_{\infty}$:

\begin{eqnarray} \label{eq:WH}
  \overline{W}^{H}=\{\overline{{\bf x}}^{H}(s)|
  \lim_{s\to-\infty}\overline{{\bf x}}^{H}(s)
   =\overline{{\bf x}}^{{\rm dht}}_{-\infty},
  \lim_{s\to \infty}\overline{{\bf x}}^{H}(s)
  =\overline{{\bf x}}^{{\rm dht}}_{ \infty} \}~,
\end{eqnarray}

\noindent
where the superscript $\{\cdot\}^{H}$ stands for
heteroclinic connection and the subscripts $\{\cdot\}_{\mp\infty}$
represent the direction of $s$ towards the corresponding DHT. In
addition, if $\overline{{\bf x}}^{{\rm dht}}_{-\infty}$ and
$\overline{{\bf x}}^{{\rm dht}}_{\infty}$ coincide, then
$\overline{W}^{H}$ is called the homoclinic connection. An
invariant manifold is special case of kinematically-defined $C$
because $s$ has a semi-infinite or bi-infinite range as in
(\ref{eq:WUS}) and (\ref{eq:WH}). We emphasize that the  terms
\emph{finite} and \emph{infinite} refer to the range of $s$ on
$C$, rather than the physical length of $C$.

For the description of the transport geometry near $C$,
it is often convenient to use an orthogonal
arc-length coordinate system,
$(l,r)$.
Along $C$,
the arc-length $l=l^{C}(s)$ and the flight-time $s$
are related by the local velocity, i.e.,
$\frac{d}{ds}l^{C}(s) =|\overline{{\bf u}}(\overline{{\bf x}}^{C} (s))|$.
Normal to $C$, $r$ is defined to be the
signed distance of a neighboring point  ${\bf x}$ to $C$;
$r>0$, $r=0$ and $r<0$ correspond to the left, on, and the right of $C$
with respect to the forward direction of $\overline{{\bf x}}^{C}(s)$
along $C$.
A pair of orthogonal unit vectors in the tangent and normal directions
to $C$ are given by

\begin{eqnarray}   \label{eq:veca}
  \mbox{\boldmath$\xi$\unboldmath}^{C}_{\parallel}(l^{C}(s))=
\frac{\overline{{\bf u}}(\overline{{\bf x}}^{C} (s))}
 {|\overline{{\bf u}}(\overline{{\bf x}}^{C}    (s))|},
 \qquad
  \mbox{\boldmath$\xi$\unboldmath}^{C}_{\perp}(l^{C}(s))=
  \frac{(-\overline{u}_{2} (\overline{{\bf x}}^{C} (s)),
  \overline{u}_{1} (\overline{{\bf x}}^{C} (s)))^{T}}
 {|\overline{{\bf u}}(\overline{{\bf x}}^{C} (s))|}~,
\end{eqnarray}

\noindent
where ${\bf u}=(u_{1},u_{2})^{T}$. The transformation
between the Cartesian and arc-length coordinates is
area-preserving.

\subsection{Unsteady flow and perturbation theory}
\label{sec:bg_unstdy}

As we have noted, the mathematical formulation of the TIME method is based on
perturbation theory for a velocity field given by
(\ref{eq:uunstdy}) and (\ref{eq:uhat}). The necessary background
and results are given in Appendix \ref{sec:pert_just}. Trajectories of the
unsteady flow passing through ${\bf x}_{0}$ on  $C$ at $t_{0}$ are
of the following form:

\begin{eqnarray} \label{eq:xTaylor-x}
    {\bf x}(t;\overline{{\bf x}}^{C}(s_{0}),t_{0}; \varepsilon)
    &=& \overline{{\bf x}}^{C}(s_{0}-t_{0}+t)
    + \varepsilon \hat{{\bf x}}(t;s_{0},t_{0} )
    + {\cal O}(\varepsilon^2),
\end{eqnarray}

\noindent where  $\varepsilon \hat{{\bf x}}(t;s_{0},t_{0} )$ is
the leading-order displacement vector with $\hat{{\bf
x}}(t_{0};s_{0},t_{0})=0$. Computing the Taylor expansion of
(\ref{eq:uunstdy}) and the time derivative of (\ref{eq:xTaylor-x})
with respect to $\varepsilon$ gives to the following  linear
ordinary differential equation for $\hat{{\bf x}}
(t;s_{0},t_{0})$:

\begin{eqnarray} \label{eq:xunstdy}
   \frac{d}{dt} \hat{{\bf x}}(t;s_{0},t_{0})
    = D_{\bf x}{\bf u}\left(
   {\overline{{\bf x}}^{C}(s_{0}-t_{0}+t)}\right)
   \hat{{\bf x}}
+  \hat{\bf u}(\overline{{\bf x}}^{C}(s_{0}-t_{0}+t),t).
\end{eqnarray}

\noindent
Given $C=\{\overline{{\bf x}}^{C}(s)\}$,
 $\hat{{\bf x}}(t;s_{0},t_{0})$ can be obtained by solving
this linear system where the nonlinear evolution of
$\overline{{\bf x}}^{C}(s_{0}-t_{0}+t)$ provides us with the
time-dependent coefficients and the inhomogeneous term. In
Appendix \ref{sec:pert_just} we show that perturbation theory can provide
valid approximations in situations  where $C$ is defined over
finite, semi-infinite or bi-infinite time intervals.

It is worth noting here that for many of the most fruitful
perturbation theories used in dynamical systems type analyses
rarely are precise bounds available for the size of the
perturbation for which the method is applicable. Nevertheless,
this has not limited the insights they have provided in a variety
of applications. For example, the typical statements of Melnikov's
method
\cite{salam_siamjam87,greenspan_holmes_book83,guckenheimer_holmes_book83}
indicate only  that it is valid for $\varepsilon$ sufficiently
small. Another example is the well-known Kolmogorov-Arnold-Moser
(KAM) theorem \cite{arnold_book78}, which has been proven useful
in many applications despite  the fact that the bounds are
generally too strict to be practically applicable. The situation
with the KAM theorem is even worse since rarely are the hypotheses
of the theorem even verified in applications since they require
the velocity to be expressed in action-angle variables, which can
rarely be achieved. This limitation also prevents one from
obtaining any type of bound on the perturbation for which the
theorem is valid.

\section{Transport functions for TIME}
\label{sec:fnc}

Having kinematically defined the Eulerian boundary $C$ by the
reference state, we now turn our attention to transport across
$C$. There are two aspects: one is concerned with the amount of
flow property and the other is concerned with geometry of
transport. Examples of  flow properties are mass, temperature,
humidity in the atmosphere,  salinity in the oceans, and such. The
TIME functions are developed for these two aspects, first for a
finite time interval along any $C$ (Section~\ref{sec:fnc_ref}) and
then for an infinite time interval along an infinite $C$
(Section~\ref{sec:fnc_mnfld}).

\subsection{Derivation of the finite-time TIME functions}
\label{sec:fnc_ref}

\subsubsection{Accumulation of a flow property}
\label{sec:fnc_ref_penetration}

We assume that the time-dependent fluctuation in the
flow property distribution, denoted by $Q({\bf x},t)$, is also small

\begin{eqnarray} \label{eq:qunstdy}
   Q ({\bf x},t) &=& \overline{Q} ({\bf x}) + Q'({\bf x},t)~.
\end{eqnarray}

\noindent
Like ${\bf u}({\bf x},t)$ in (\ref{eq:uunstdy}) and
(\ref{eq:uhat}), we introduce $Q'({\bf x},t)=\varepsilon
\hat{Q}({\bf x},t)$ with $|Q'(\overline{{\bf x}}^{C}(s),t)| = {\cal O}
(\varepsilon|\overline{Q}(\overline{{\bf x}}^{C}(s))|)$.
To illustare the basic idea for estimating the amount of property
transport, we consider the imaginary  fluid column ${\cal F}$
in the flow (Figure~\ref{fg:string}c).
By accumulating the flux at the moving intersection of ${\cal F}$
with $C$, we obtain the net amount of accumulation.

\fginsrt{fg:string}

Up to leading order, the intersection $(s,t)$ of ${\cal F}$ with
$C$  at time $t$ is approximated by the reference trajectory
$(s_{0}-t_{0}+t,t)$ using perturbation theory (Appendix~\ref{sec:P}). 
At $(s,t)$, the
instantaneous flux of $Q(\overline{{\bf x}}^{C}(s),t)$ carried by
the local velocity ${\bf u}(\overline{{\bf x}}^{C}(s),t)$ across
$C$ per unit length is

\begin{eqnarray}
 &&  \mbox{\boldmath$\xi$\unboldmath}^{C}_{\perp}(l^{C}(s))~\cdot ~
   [~Q(\overline{{\bf x}}^{C}(s),t)~{\bf u}(\overline{{\bf x}}^{C}(s),t)~]
 = \nonumber \\
 &&  \mbox{\boldmath$\xi$\unboldmath}^{C}_{\parallel}(l^{C}(s))\wedge
   [~\overline{Q}(\overline{{\bf x}}^{C}(s))
    ~\varepsilon \hat{{\bf u}} (\overline{{\bf x}}^{C}(s),t) +
   \varepsilon^{2}
 \hat{Q}(\overline{{\bf x}}^{C}(s),t)~
 \hat{{\bf u}}(\overline{{\bf x}}^{C}(s),t) ~]~,
\label{eq:qu}
 \end{eqnarray}

 \noindent
where $\mbox{\boldmath$\xi$\unboldmath}^{C}_{\perp}(l^{C}(s))$ and
$\mbox{\boldmath$\xi$\unboldmath}^{C}_{\parallel}(l^{C}(s))$
are defined in (\ref{eq:veca}).
The positive value means the flux from the right to
the left across $C$ with respect to the forward direction of $s$.
This formula (\ref{eq:qu}) says that
 the instantaneous flux of $Q$
across $C$ exists if  ${\bf u}({\bf x},t)$ has
a component normal to $C$, and that the
time-dependent fluctuation
$Q'({\bf x},t)=\varepsilon \hat{Q}({\bf x},t)$
contributes to the transport at the higher order.
At the  leading order, the instantaneous flux of $Q$
penetrating across $C$ at $(s,t)$ per unit flight time   is
$\overline{Q}(\overline{{\bf x}}^{C}(s))~ 
\overline{{\bf u}}(\overline{{\bf x}}^{C}(s)) \wedge
\varepsilon \hat{{\bf u}} (\overline{{\bf x}}^{C}(s),t) $, i.e.,

 \begin{subequations} \label{eq:Qinst}
  \begin{eqnarray}    \label{eq:Qinst-Q}
  \overline{Q}(\overline{{\bf x}}^{C}(s))~
\overline{{\bf u}}(\overline{{\bf x}}^{C}(s)) \wedge
   {\bf u}'(\overline{{\bf x}}^{C}(s),t)
 &=& \overline{q}^{C} (s)~\mu^{C}(s,t)
  \end{eqnarray}
using (\ref{eq:uhat}), where
  \begin{eqnarray}    \label{eq:pinst}
    \overline{q}^{C} (s) &\equiv& \overline{Q} (\overline{{\bf x}}^{C}(s)) \label{eq:Qinst-q}\\
    \mu^{C}(s,t)&\equiv&\overline{{\bf u}}(\overline{{\bf x}}^{C}(s))\wedge{\bf u}'(\overline{{\bf x}}^{C}(s),t).
    \label{eq:muinst}
  \end{eqnarray}
 \end{subequations}

\noindent  We refer to $\mu^{C}(s,t)$ as the \emph{instantaneous flux
  function}, induced by the  unsteadiness
  (eddy) of the velocity  through the interaction with
  the reference (mean) flow.
  This is the origin of the transport induced by the mean-eddy
 interaction (TIME) across $C$.
  The sign of $\mu^{C}(s,t)$ indicates the direction of the
  instantaneous flux.

The accumulation over the interval $[t_{0},t_{1}]$  is
thus approximated  by
$ \int_{t_{0}}^{t_{1}}\overline{q}^{C} (s_{0}-t_{0}+\tau) ~
 \mu^{C}(s_{0}-t_{0}+\tau,\tau) d\tau$
up to leading order.
Because this amount is the same for any $(s,t)$ along the reference
trajectory with  $s-t=s_{0}-t_{0}$,
we obtain a general form of the accumulation
\begin{eqnarray}    \label{eq:mint}
   m^{C} (s,t;t_{0}:t_{1})&\equiv&
  \int_{t_{0}}^{t_{1}} \overline{q}^{C} (s-t+\tau)
    ~ \mu^{C}(s-t+\tau,\tau) d\tau~,
\end{eqnarray}
 where  the first pair $(s,t)$ in the arguments
of the left-hand side represents the combination of spatial coordinate
and time at which the net accumulation of $Q$ is evaluated,
while the next pair $(t_{0}:t_{1})$ concerns the time interval on
which the transport takes place.
Here $t$  can be either before or after $t_{0}$ or $t_{1}$.
We refer to $m^{C}(s,t;t_{0}:t_{1})$ as the
\emph{accumulation  function}.
Characteristics of $m^{C}(s,t;t_{0}:t_{1})$
will be discussed further in  Section~\ref{sec:chr}.

\subsubsection{Displacement distance and area}
\label{sec:fnc_ref_displacement}

For the geometry, we consider the displacement
 distance of the particle starting from $(s_{0},t_{0})$ on $C$.
In the unsteady flow at time $t$, the displacement of
the particle from $\overline{{\bf x}}^{C}(s_{0}-t_{0}+t)$
is ${\bf x}'(t;\overline{{\bf x}}^{C}(s_{0}),t_{0})$
up to  leading order by (\ref{eq:xunstdy}).
For particle transport and its geometry,
we choose to  use arc-length coordinates $(l,r)$
in the description of the displacement functions
because the displacement distance has the physical dimension of length.
Using (\ref{eq:xTaylor-x}) along a reference trajectory
$(s,t)=(s_{0}-t_{0}+t)$ with the initial condition
$(l_{0},t_{0})=(l^{C}(s_{0}),t_{0})$,
 the leading order term for the displacement distance
 due to particle transport at $t\in[t_{0},t_{1}]$ is given by

\begin{subequations} \label{eq:rao}
   \begin{eqnarray}
    r'(t;l_{0},t_{0})&\equiv&
    \mbox{\boldmath$\xi$\unboldmath}^{C}_{\parallel}
    (l^{C}(s_{0}-t_{0}+\tau))
     \wedge{\bf x}'(t;s_{0},t_{0}) ~.
   \label{eq:rao-r}
   \end{eqnarray}

   \noindent
Because ${\bf x}'(t;s_{0},t_{0})
=\varepsilon\hat{{\bf x}}(t;s_{0},t_{0})$ can be obtained by solving
 (\ref{eq:xunstdy}), then so can $r'(\tau;l_{0},t_{0})$
 by the direct substitution.
However,
a simpler formula is available by considering

   \begin{eqnarray}
    a'(t;l_{0},t_{0})&\equiv&
  r'(t;l_{0},t_{0})|\overline{{\bf u}}(\overline{{\bf x}}^{C}(s))|~=
    \overline{{\bf u}}(\overline{{\bf x}}^{C}(s_{0}-t_{0}+t))
  \wedge{\bf x}'(t;s_{0},t_{0})~, \nonumber \\
      \label{eq:rao-a}
   \end{eqnarray}
 \end{subequations}

 \noindent
which corresponds to the displacement area per unit $s$ along $C$
as shown in  Appendix~\ref{sec:pert_just}.
 Using $a'(t_{0};l_{0},t_{0})=0$ for the initial condition,
 construction, the solution for the displacement area at $t_{1}$ is
 given by

  \begin{eqnarray}\label{eq:ao-a}
    a'(t_{1};l_{0},t_{0}) & = & \int_{t_{0}}^{t_{1}} \overline{e}^{C}
    (s-t+t_{1}:s-t+\tau)\
    \mu^{C}(s-t+\tau,\tau) d\tau~,
  \end{eqnarray}

  \noindent
 where

  \begin{eqnarray}\label{eq:ao-e}
   \overline{e}^{C} (s:s_{0})&\equiv& \exp\{\int_{s_{0}}^{s} {\rm trace}
     \{D_{\bf x}\overline{{\bf u}}(\overline{{\bf x}}^{C}(\theta)) \} d\theta\}
  \end{eqnarray}
 reflects the compressibility of the reference flow;
 for an incompressible flow, $\overline{e}^{C} (s:s_{0})\equiv 1$.

Like the accumulation,
the displacement area associated with transport over
$[t_{0},t_{1}]$ can be evaluated at $(l^{C} (s),t)$
where $t$ can be before, in, or after the time interval.
Conceptually, this is to let
 $a'(t_{1};l_{0},t_{0})$ obtained by (\ref{eq:ao-a}) evolve
 under the reference flow over an additional time interval
 $[t_{1},t]$ to take the incompressibility into account.
As shown in Appendix~\ref{sec:pert_just},
the final form of the displacement functions is  given by

  \begin{subequations}\label{eq:ra}
  \begin{eqnarray}
  a^{C}(s,t;t_{0}:t_{1}) &=&\overline{e}^{C} (s:s_{0}-t_{0}+t_{1}) a'(t_{1};l_{0},t_{0})
  \nonumber \\
  &=& \int_{t_{0}}^{t_{1}}\overline{e}^{C} (s:s-t+\tau)\
  \mu^{C}(s-t+\tau,\tau) d\tau, \nonumber \\ \label{eq:ra-a} \\
  r^{C}(l^{C} (s),t;t_{0}:t_{1}) &=&
   \frac{a^{C}(s,t;t_{0}:t_{1})}
   {|\overline{{\bf u}}(\overline{{\bf x}}^{C}(s))|}~.
    \label{eq:ra-r}
  \end{eqnarray}
\end{subequations}
As in the case of $m^{C}(s,t;t_{0}:t_{1})$,
the first pair $(l^{C} (s),t)$
in the argument represents the spatial coordinate and time at which the
function is evaluated, and the next pair correspond to the time interval
when transport takes place.

 Accordingly over $[t_{1},t_{0}]$, the displacement is determined
 by two contributions: one is from the unsteadiness of the flow measured
 along $C$ through the instantaneous flux $\mu^{C}(s-t+\tau,\tau)$,
 and the other from the compressibility of the reference flow
 through  $\overline{e}^{C}(s:s-t+\tau)$, which may result in compression or
 expansion of  the  area.
Sign of
 $a^{C}(s,t;t_{0}:t_{1})$ indicates the
directionality of transport across $C$.

\subsection{Extension over the infinite TIME functions}
\label{sec:fnc_mnfld}

We refer to the accumulation function (\ref{eq:mint}) and
displacement distance and displacement area functions
(\ref{eq:ra}) as the \emph{(finite) TIME functions} because they are defined
over a finite time interval and hence the finite range of $s$ along $C$.
These TIME functions can be extended over
the semi-infinite $\overline{W}^{U}$ and $\overline{W}^{S}$ and bi-infinite
$\overline{W}^{H}$, because the exponential decay of the velocity towards
DHTs at the starting or(and) end point(s) guarantees the convergence
conditions required for the validity of perturbation theory
(Appendix~\ref{sec:pert_just}).
The extension of the displacement functions over
$\overline{W}^{H}$ can be particularly useful since it provides a direct
link to the Melnikov function which measures the leading
order distance between the \emph{time-dependent} unstable and stable
invariant manifolds
\cite{salam_siamjam87,greenspan_holmes_book83,guckenheimer_holmes_book83}.
The Melnikov function has been  used to study Lagrangian
transport, mostly in incompressible flows
\cite{romkedar_etal_jfm90}; also see \cite{samwig}.

The TIME functions for these special $C$ are as follows.
The transport that has happened in the past
across $\overline{W}^{U}$
can be obtained by extending the TIME functions over
a semi-infinite time interval $(-\infty,t_{0}]$:
  \begin{subequations} \label{eq:mrapst}
   \begin{eqnarray}
    m^{U} (s,t;t_{0})
     &=&\int_{-\infty}^{t_{0}} \overline{q}^{C} (s-t+\tau) ~
     \mu^{C}(s-t+\tau,\tau)d\tau;
     \label{eq:mrapst-m} \\
    a^{U}(s,t;t_{0})
    &=& \int_{-\infty}^{t_{0}}\overline{e}^{C}
     (s:s-t+\tau) ~ \mu^{C}(s-t+\tau,\tau)d\tau;
     \label{eq:mrapst-a} \\
    r^{U}(l^{U}(s),t;t_{0}) &\equiv&
     \frac{a^{U}(s,t;t_{0})} {|\overline{{\bf u}}(\overline{{\bf x}}^{U}(s))|}.
     \label{eq:mrapst-r}
  \end{eqnarray}
for some $t_{0}$.
 \end{subequations}
\begin{subequations} \label{eq:mraftr}
 Similarly,  the  transport that will happen in the future
 across  the stable manifold $\overline{W}^{S}$
 can be obtained by extending the TIME functions
 over  the semi-infinite interval $[t_{0},\infty)$:
  \begin{eqnarray}
  m^{S}(s,t;t_{0})
  &=& \int_{t_{0}}^{\infty} \overline{q}^{C} (s-t+\tau) ~
 \mu^{C}(s-t+\tau,\tau) d\tau;  \label{eq:mraftr-m} \\
 a^{S}(s,t;t_{0})
 &=& \int_{t_{0}}^{\infty}\overline{e}^{C} (s:s-t+\tau) ~
 \mu^{C}(s-t+\tau,\tau)d\tau; \label{eq:mraftr-a} \\
 r^{S}(l^{S}(s),t;t_{0}) &\equiv& \frac{a^{S}(s,t;t_{0})}
 {|\overline{{\bf u}}(\overline{{\bf x}}^{S}(s))|}.
     \label{eq:mraftr-r}
 \end{eqnarray}
 \end{subequations}
 Finally,  the entire transport across  the separatrix
 $\overline{W}^{H}$ can be obtained by extending the TIME
 functions over the bi-infinite interval $(-\infty,\infty)$:
\begin{subequations} \label{eq:mrainf}
 \begin{eqnarray}
  m^{H} (s,t)&=&m^{U}(s,t;t_{0})+m^{S}(s,t;t_{0}) \nonumber \\
 &=&  \int_{-\infty}^{\infty} \overline{q}^{C} (s-t+\tau) ~
  \mu^{C}(s-t+\tau,\tau) d\tau; \label{eq:mrainf-m} \\
 a^{H}(s,t) &=&
  a^{U}(s,t;t_{0})+a^{S}(s,t;t_{0}) \nonumber \\
 &=&  \int_{-\infty}^{\infty}\overline{e}^{C} (s:s-t+\tau) ~ \mu^{C}
  (s-t+\tau,\tau)d\tau;     \label{eq:mrainf-a} \\
    r^{H}(l^{H}(s),t) &\equiv&
     r^{U}(l^{H}(s),t;t_{0})+r^{S}(l^{H}(s),t;t_{0}) =
     \frac{a^{H}(s,t;t_{0})}{|\overline{{\bf u}}(\overline{{\bf x}}^{H}(s))|}.
     \label{eq:mrainf-r}
 \end{eqnarray}
\end{subequations}
The displacement area function $a^{H}(l^{H}(s),t)$ for $\overline{W}^{H}$ is
the same as the so-called Melnikov function.

\section{Characteristics of TIME}
\label{sec:chr}

\subsection{Characteristics along an individual trajectory}
\label{sec:chr_individual}

The accumulation is obtained by following the
individual reference trajectories  (Section~\ref{sec:fnc}). 
This leads to the
concepts of \emph{invriance} and \emph{piece-wise independence}.

\noindent
\textsf{Invariance of the accumulation function.}
Given a fixed time interval $[t_{0},t_{1}]$, the accumulation function
is \emph{invariant}:
 \begin{eqnarray} \label{eq:mdt}
 m^{C}(s,t;t_{0}:t_{1}) & = &
 m^{C}(s-t+\tau,\tau;t_{0}:t_{1}) 
 \label{eq:mdt-p} 
 \end{eqnarray}
for any $\tau$. 
This invariance implies that each trajectory has perfect memory for the
 amount of transport.
Invariance for the displacement area function:
 \begin{eqnarray} \label{eq:adt}
 a^{C}(s,t;t_{0}:t_{1}) & = &
  \overline{e}^{C}  (s:s-t+\tau)~
        a^{C}(s-t+\tau,\tau;t_{0}:t_{1})~
 \label{eq:adt-p}
 \end{eqnarray}
is subject to the compressibility factor 
$\overline{e}^{C}  (s:s-t+\tau)$ of the reference flow.

\textsf{Piece-wise independence.}
It is clear in the definition (\ref{eq:mint})
that the time interval $[t_{0},t_{1}]$ can be broken 
up into an arbitrary number (say $K$) of pieces:
 \begin{eqnarray} \label{eq:msum}
 m^{C}(s,t; t_{0}:t_{1}) & = &
 \sum_{k=1}^{K} m^{C}(s,t; t_{0.k-1}:t_{0.k})~,
    \label{eq:msum-mc}
 \end{eqnarray}
where
$\cup_{k=1}^{K}[t_{0.k-1},t_{0.k}]=[t_{0},t_{1}]$
with $t_{0.0}=t_{0}$ and $t_{0.K}=t_{1}$.
This is a \emph{temporal piece-wise independence}.
Using (\ref{eq:xmn-sgm}) along $(s_{0}-t_{0}+\tau,\tau)$,
\emph{spatial piece-wise independence} follows naturally by
breaking the spatial segment
$[s\sb{a},s\sb{b}]$ into $K$ pieces by
$\cup_{k=1}^{K}[s_{a.k-1},s_{a.k}]=[s\sb{a},s\sb{b}]$
with $s_{a.0}=s\sb{a}$ and $s_{a.K}=s\sb{b}$
and transforming them into $K$ temporal pieces 
with $[t\sb{0.k-1}, t\sb{0.k}] =
 [t_{0}-s\sb{a.0}+s\sb{a.k-1},t_{0}-s\sb{a.0}+s\sb{a.k}]$ 
over $[t_{0},t_{1}]=[t_{0},t_{0}-s\sb{a}+s\sb{b}]$.
Because the independence is a characteristics defined 
for a fixed $(s,t)$,
 both temporal independence and spatial independence
hold for the displacement area function $a^{C}(s,t;t_{0}:t_{1})$.

\subsection{Coherency of transport}
\label{sec:chr_geo} 

Geometry of particle displacement 
leads to the  concepts associated with the coherency of transport.
Because the geometrical
characteristics discussed here hold for any $C$ over any time
interval, we will drop $[t_{0}:t_{1}]$ from each notation for
simplicity after the first appearance as indicated in  
$[\equiv\{\cdot\}]$; for example, 
$r^{C}(l^{C}(s),t)[\equiv r^{C}(l^{C}(s),t;t_{0}:t_{1})]$. 
When $C$ is taken as $\overline{W}^{H}$,
there is a geometrical relation to Lagrangian transport, which we
will treat separately in Appendix~\ref{sec:L}.

\begin{figure}[htb!]
\begin{center}
\includegraphics[height=6.cm]{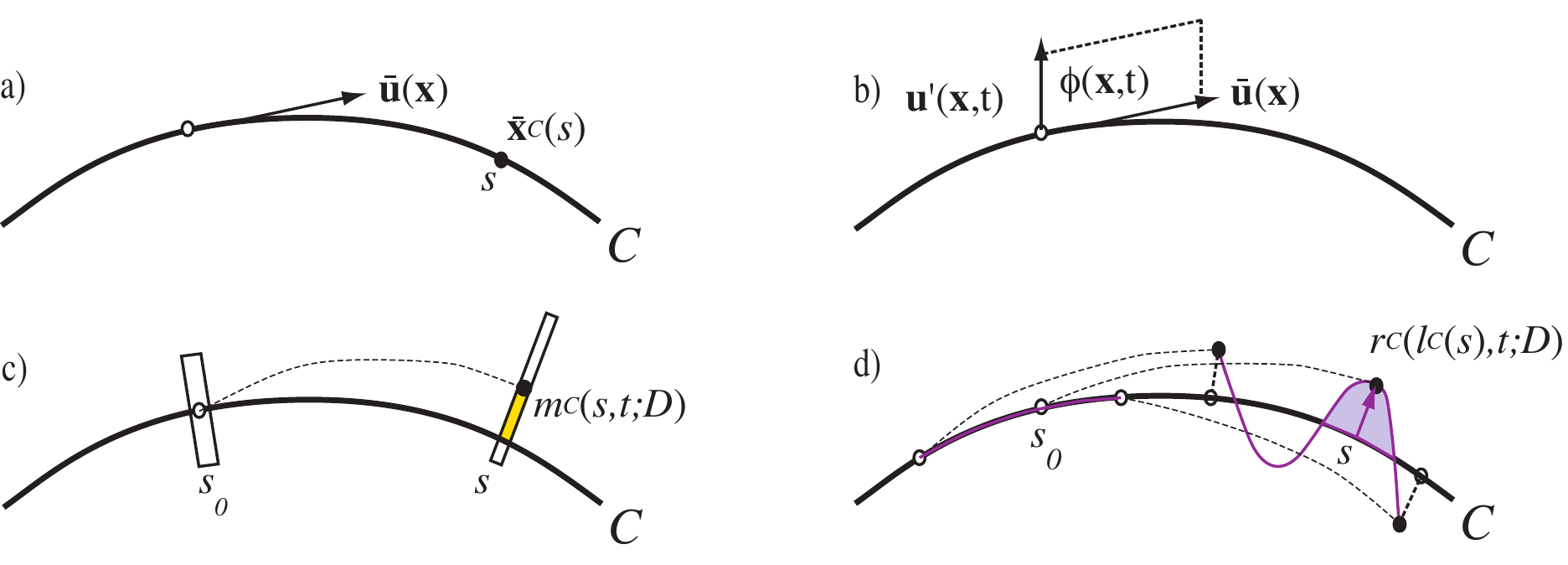}
\end{center}
\caption{Geometry associated with the TIME method:
a) Eulerian boundary $C=\{ \overline{{\bf x}}^{C}(s)\}$ and the mean velocity $\o
verline{{\bf u}}({\bf x})$;
b) instantaneous flux $\phi({\bf x},t)$ across $C$ 
 as the parallelogram in the unsteady flow;
c) accumulation $m^{C}(s,t;D)$ 
   (the shaded portion represents the accumulation); 
d) displacement distance $m^{C}(s,t;D)$ 
   (the shaded area represents a positive pseudo-lobe area).}
\label{fg:string}
\end{figure}

For an illustration of transport geometry and coherency,
let us consider an imaginary material curve ${\cal R}(t)$ placed 
initially on $C$ at  time $t_{0}$, i.e., 
${\cal R}(t_{0})=\{(l,r)|r=0\}$ (Figure~\ref{fg:string}d).
In the reference flow, ${\cal R}(t)$ advects
along $C$ without any displacement. 
In the  unsteady flow, velocity
normal to $C$ may let ${\cal R}(t)$ depart from $C$.
For the transport geometry associated with the TIME method,
we define
\begin{eqnarray} \label{eq:Rt}
R^{C}(t) &=& \{(l,r) | l=l^{C}(s),r=r^{C}(l^{C}(s),t)\}~,
\end{eqnarray}
which is the leading order  approximation
to ${\cal R}(t)$.

\textsf{Pseudo-primary intersection point (pseudo-PIP)
 sequence of $R^{C}(t)$ and $C$.}
In the unsteady flow, $R^{C}(t)$ may intersect with
$C$ to form a chain of lobe-like structures 
as shown schematically in Figure~\ref{fg:string}d.
We denote such a ordered sequence of such zeros
$\{s^{C}_{j}(t)[\equiv s^{C}_{j}(t;t_{0}:t_{1})]\}$ with 
\begin{eqnarray}\label{eq:sCjt}
s^{C}_{j}(t)
&=&\{s~|~a^{C}(s,t)=0, ~s^{C}_{j}(t)<s^{C}_{j+1}(t)\}~,
\end{eqnarray}
where we use the fact that the zeros of 
$a^{C}(s,t)[\equiv a^{C}(s,t;t_{0}:t_{1})]$ and those of
$r^{C}(l,t)$ are identical.
Unless the zero is non-degenerate,
$\{s^{C}_{j}(t)\}$ is generally identical to the intersection sequence of
 ${\cal R}(t)$ up to the leading order;
see \cite{wiggins_book92} and also Appendix~\ref{sec:P}.
We call $\{s^{C}_{j}(t)\}$ the
\emph{pseudo-primary intersection point (pseudo-PIP)} sequence
in contrast to the Lagrangian lobe dynamics 
for the heteroclinic connection.
The term 'primary' is  used here to emphasize the analogy of the
PIP of Lagrangian lobe counterpart.
A pseudo-PIP sequence can be transformed into arc-length coordinate
$l^{C}_{j}(t)[\equiv l^{C}_{j}(t;t_{0}:t_{1})]$ so that
$l^{C}\sb{j}(t)\equiv l^{C}(s^{C}_{j}(t))$.
%
\fginsrt{fg:Elobe}

\textsf{Invariance of pseudo-PIP sequence.}
 Because the displacement area 
 $a^{C}(s,t)$
 is invariant
 subject to the  compressibility effect  from (\ref{eq:adt}) and  the
 compressibility effect will not change  $\{s^{C}_{j}(t)\}$ at given $t$,
 $\{s^{C}_{j}(t)\}$  is invariant. 
Each $s^{C}_{j}(t)$ coincide with a reference 
 trajectory, i.e.,
 \begin{eqnarray} \label{eq:rz-s}
   (s^{C}_{j}(\tau),\tau) &=&(s^{C}_{j}(t)-t+\tau,\tau).
 \end{eqnarray}

\textsf{Pseudo-lobe seqence.}
 We denote the lobe-like structure defined by the
 segments of $R^{C} (t)$ and $C$ between a pair of two adjacent 
 pseudo-PIPs 
 by $L^{C}_{j,j+1}(t)[\equiv  L^{C}_{j,j+1}(t;t_{0}:t_{1})]$
 and call it the \emph{pseudo-lobe}:
\begin{eqnarray}\label{eq:LCjjt}
L^{C}_{j,j+1}(t) &=&
\{(l,r)~|~ r(r-r^{C}(l,t))\leq 0,~l\in[l^{C}_{j}(t),l^{C}_{j+1}(t)]\}~.
\end{eqnarray}
 Its sequence makes the chain-like structure, which we call the
 \emph{pseudo-lobe sqeuence}, $\{L^{C}_{j,j+1}(t)\}$.
 Using $a^{C}(s,t)$, the pseudo-lobe is the area surrounded by 
 $a^{C}(s,t)$ and $C$ between two adjacent pseudo-PIPs.
 It is worth emphasizing that temporal and spatial piece-wise
 independence (Section~\ref{sec:chr_individual}) 
 holds for the displacement function of each pseudo-lobe individually.

\textsf{Signed area.}
 The size of each  pseudo-lobe measures the amount of the locally coherent
 transport.
 Its \emph{signed area} is given by:
 \begin{eqnarray}
   A( L^{C}_{j,j+1}(t) )
   & = &  \int_{l^{C}_{j}(t)}^{l^{C}_{j+1}(t)}r^{C}(l,t) dl
   = \int_{s^{C}_{j}(t)}^{s^{C}_{j+1}(t)}a^{C}(s,t) ds. \label{eq:AL}
 \end{eqnarray}
 If the flow is incompressible, the area of each
 lobe is invariant, i.e.,
 $A( L^{C}_{j,j+1}(\tau) )=A( L^{C}_{j,j+1}(t) )$.

\textsf{Directionality of pseudo-lobes.}
 Each pseudo-lobe represents the amount of
 fluid particles that go across $C$ over  $[t_{0},t_{1}]$.
 Depending on whether $R^{C}(t)$ lies to the left or right of $C$, 
 $ L^{C}_{j,j+1}(t)$ can be  of two types, 
 denoted respectively by $L^{R C}$ or $L^{C R}$ corresponding to   
 $r^{C}(l,t)>0$ or $r^{C}(l,t)<0$
 for transport across $C$ from right to left  or left to right.
 Provided all intersections are transverse,
 $\{L^{C}_{j,j+1}(t) \}$ alternated the types between 
 $L^{R C}$ and $L^{C R}$ along $C$, resulting in  a chain structure
 at any given time. 


The geometry associated with transport of $Q$ can be established 
conceptually by replacing the displacement area function
$a^{C}(s,t)$ for $R^{C}(t)$ with the
accumulation function
$m^{C}(s,t)[\equiv m^{C}(s,t;t_{0}:t_{1})]$ in arc-length coordinates.
In the case  for mass
with $\overline{q}^{C}(s)\equiv 1$
where the flow is incompressible, this
operation results exactly in the displacement functions of
 fluid particles, $m^{C}(s,t)=a^{C}(s,t)$.

\section{Application to a numerical simulation of the
wind-driven double-gyre ocean circulation}
\label{sec:application}

\begin{figure}[!ht]
\begin{center}
\includegraphics[height=12.cm]{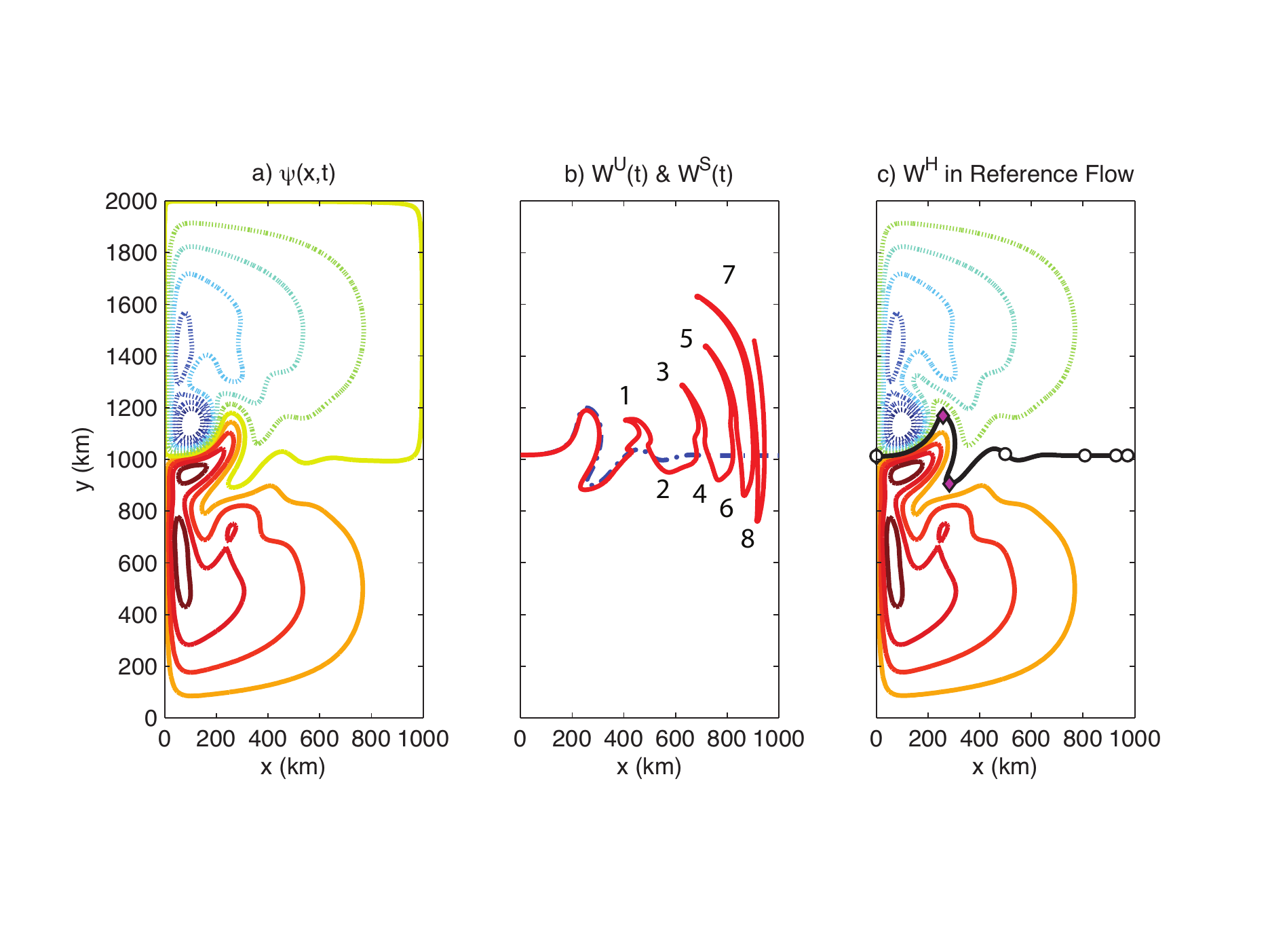}
\end{center}
\caption{Double-gyre circulation:
 a) streamfunction $\psi({\bf x},t)$ at $t=t^{*}_{944}$;
 b) unstable manifold ${\cal W}^{U}(t)$ (solid line) and 
 stable manifold ${\cal W}^{s}(t)$ (dashed line) of the inter-gyre
 transport by the Lagrangian method; and
 c) reference streamfunction $\overline{\psi}({\bf x})$
  and the Eulerian boundary 
 $\overline{W}^{H}=\overline{{\bf x}}^{H}(s)$ of the inter-gyre
 transport by the TIME method.
 In (a) and (c), contour interval 2000 with dash lines 
 for negative values.
 In (b), numbers correspond to those in Table~\ref{tbl:dg_EL}
 In (c),  $\overline{W}^{H}=\overline{{\bf x}}^{H}(s)$  is the 
 thick solid line with four white circles plotted at
 every 250 days starting from in $s=0$, i.e.,
 $\overline{{\bf x}}^{H}(s^{*}_{0})$,
 $\overline{{\bf x}}^{H}(s^{*}_{250})$,
 $\overline{{\bf x}}^{H}(s^{*}_{500})$, and 
 $\overline{{\bf x}}^{H}(s^{*}_{1000})$;
 the diamonds  at
  $\overline{{\bf x}}^{H}(s_{{\rm J}})$,
  $\overline{{\bf x}}^{H}(s_{{\rm N}})$, and 
  $\overline{{\bf x}}^{H}(s_{{\rm S}})$, with
   $(s_{{\rm J}},s_{{\rm N}},s_{{\rm S}})
  =(s^{*}_{110},s^{*}_{129},s^{*}_{174.5})$;
 $\overline{{\bf x}}^{H}(s_{{\rm J}})$ almost overlaps
    $\overline{{\bf x}}^{H}(s^{*}_{0})$.
 }
\label{fg:dg_psi_mnf_C}
\end{figure}

In this study we apply the TIME method to the inter-gyre transport
in the mid-latitude, wind-driven ocean circulation.
The data set is
obtained by a numerical simulation of a quasi-geostrophic (QG)
3-layer model  in a rectangular basin geometry with free slip
boundary conditions \cite{rowley_pd96}. 
Due to the latitudinal
antisymmetric wind-stress curl applied at the ocean surface, the
basic circulation pattern in the top layer is a double-gyre
structure separated by an eastward jet shooting off from the
confluence point of the southward  and northward western boundary
currents (Figure~\ref{fg:dg_psi_mnf_C}). 
Driven by this strong jet, 
the subpolar gyre circulates counterclockwise in the north and 
the subtropical gyre circulates clockwise in the south.
Depending on the value of the parameters such as the
viscosity and the wind stress curl,
the ocean circulation exhibits a rich time-dependent dynamics
\cite{dijkstra_book05,simonnet_etal_jpo03}.
\fginsrt{fg:dg_psi_mnf_C}

At a wind stress curl of  $0.165 ~{\rm dyn}/{\rm cm}^{2}$, 
the ocean dynamics is nearly periodic with dominant spectral peak 
at  period $T\approx 151$days 
in a 1000 km $\times$ 2000 km rectangular domain. 
We choose this flow since the physical interpretation for a flow 
field close to periodic is more simple and therefore it allows us to
focus more on demonstrating our method. 
It is worth noting again that the method itself does not require  time
periodity of the flow.
The velocity data set used in this study 
has a spatial  resolution of 12.5km$\times$12.5km 
and is saved daily after a 30,000-day spin-up from  rest. 
Figure~\ref{fg:dg_psi_mnf_C}a shows the streamfunction $\psi({\bf x},t)$ of
the top layer at $t=t^{*}_{944}$ when  $\psi({\bf x},t)$ is close to
the reference state $\overline{\psi}({\bf x})$.
From here on, the subscript in $t^{*}$ denotes days after the   completion of the
spin-up. 
The most significant region of unsteadiness of the flow lies in the upstream
region of the eastward jet near the western boundary, while 
small-amplitude Rossby waves propagate westward in the entire ocean
basin.
The Lagrangian transport processes
between the two gyres are governed by the lobe dynamics 
associated with the unstable invariant manifold ${\cal W}^{U}(t)$ 
of the upstream DHT 
on  the western boundary 
and the stable invariant manifold ${\cal W}^{S}(t)$ 
of the downstream DHT
on  the eastern boundary;
see  Appendices~\ref{sec:P} and~\ref{sec:L} for the definitions 
and more details of  ${\cal W}^{U}(t)$  and ${\cal W}^{S}(t)$. 
The inter-gyre transport  in the top layer was carefully studied
by \cite{coulliette_wiggins_npg00} using  Lagrangian lobe dynamics methods.

For the comparison with the Lagrangian method,
we choose the Eulerian boundary of the TIME method
as the bi-infinite, reference heteroclinic connection 
$\overline{W}^{H}=\{\overline{{\bf x}}^{H}(s)\}$ which
 spans over $s\in(-\infty,\infty)$ with 
$\overline{{\bf x}}^{H}(-\infty)=\overline{{\bf x}}^{{\rm dht}}_{-\infty}$ 
on the western boundary
and $\overline{{\bf x}}^{H}(\infty)=\overline{{\bf x}}^{{\rm dht}}_{+\infty}$ 
on the eastern
boundary (solid line in Figure~\ref{fg:dg_psi_mnf_C}c).
It is worth noting again  that application of the TIME method is not limited
to flows that possess a heteroclinic connection. 
For the computation of the TIME functions,
we choose the location of $s=0$ so that $\overline{{\bf x}}^{H}(0)$ 
is very close to $\overline{{\bf x}}^{{\rm dht}}_{-\infty}$ and
$|\overline{{\bf u}}(\overline{{\bf x}}^{H}(0))|$ 
is exponentially close to zero.
Along $\overline{W}^{H}$, the speed 
$|\overline{{\bf u}}(\overline{{\bf x}}^{H}(s))|$ of the mean jet
significantly increases starting near 
$\overline{{\bf x}}^{H}(s_{{\rm J}})$.
Geographically $\overline{{\bf x}}^{H}(s_{{\rm J}})$ is separated from 
$\overline{{\bf x}}^{H}(0)$  only by 600m.
As $s$ increases towards the downstream direction,  $\overline{W}^{H}$  makes
a sharp turn in around $\overline{{\bf x}}^{H}(s_{{\rm N}})$ in the
north, followed by the second sharp turn
around $\overline{{\bf x}}^{H}(s_{{\rm S}})$ in the south. 
Measured in the flight time,
$s_{{\rm J}}$, $s_{{\rm N}}$, and $s_{{\rm S}}$ are 110days,
129days, and 174.5days, respectively.
In the further downstream direction,  
$\overline{W}^{H}$ extends to the east and exhibits little meandering.
After about $\overline{{\bf x}}^{H}(s^{*}_{250})$
with the subscript in $s^{*}$  for the
flight-time coordinate in days from $s=0$, 
$|\overline{{\bf u}}(\overline{{\bf x}}^{H}(s))|$
becomes extremely small.
In the region near $\overline{W}^{H}$,  the order of the unsteadiness
relative to the reference state is small ($\sim0.1$),
supporting the applicability of the TIME method 
(see Appendix~\ref{sec:P_data}).

Figure~\ref{fg:dg_must} shows the instantaneous flux $\mu^{H}(s,t)$ 
on $\overline{W}^{H}$ 
as a Hovm\"{o}ller diagram \cite{hovmoller_tellus49,martis_etal_tellus06}
in the $(s,t)$ space for $[0,500]\times[750,1250]$.
\fginsrt{fg:dg_must}
The signals of $\mu^{H}(s,t)$ are periodic in $t$ with the period $T$ 
because of the time-periodic ocean dynamics.
For small $s$,
$\mu^{H}(s,t)$ is always near zero because
$\overline{{\bf u}}(\overline{{\bf x}}^{H}(s))$ is exponentially small 
near $\overline{{\bf x}}^{{\rm dht}}_{-\infty}$.
Also for large $s$, $\mu^{H}(s,t)$ is always almost zero
because both $\overline{{\bf u}}(\overline{{\bf x}}^{H}(s))$ and 
${\bf u}'(\overline{{\bf x}}^{H}(s),t)$ are small there.
Most signals of $\mu^{H}(s,t)$ are confined in between.
Figure~\ref{fg:dg_mus} shows the four
different phases of $\mu^{H}(s,t)$
during one period of the ocean oscillation at
$t=$ $t^{*}_{944}$, $t^{*}_{982}(=t^{*}_{944}+T/4)$, 
$t^{*}_{1020}(=t^{*}_{944}+T/2)$,  
and $t^{*}_{1058}(=t^{*}_{944}+3T/4)$.
\fginsrt{fg:dg_mus}
Downstream propagation of the large signals 
is seen over $[s_{{\rm J}},s_{{\rm N}}]$ where
typically three extremes 
(one local minimum and two local maxima, or {\it vice versa}, 
depending on the phase of the ocean dynamics during $T$)
propagate downstream.
Over the subsequent segment,
upstream propagation of the small signals is observed.
These signals are a consequence of the  double-gyre ocean dynamics 
(variability) through the interaction with the reference (mean) flow.
In the companion paper \cite{ide_wiggins_pd06b}, 
we examine the details of the variability in the double-gyre ocean
dynamics and its relation to transport processes, while
the focus of this paper is on the comparison of the results of the TIME method
with  those obtained by the Lagrangian lobe dynamics method.
For the computation of the TIME functions,
it suffices to use $\mu^{H}(s,t)$ 
over a short finite segment rather than the entire $(-\infty,\infty)$
 along $\overline{W}^{H}$ because $\mu^{H}(s,t)$ decays to zero for both
 small and large $s$.
Our results presented here use $\mu^{H}(s,t)$  over 
$[s^{*}_{0},s^{*}_{1000}]$.
A straight diagonal line in Figure~\ref{fg:dg_must} is an example of  the
reference trajectory  
along which the TIME functions are integrated.

Figure~\ref{fg:dg_ms} shows the infinite displacement area functions.
\fginsrt{fg:dg_ms}
By choosing $t_{0}=t$ in (\ref{eq:mrapst})-(\ref{eq:mrainf}),
$a^{U}(s,t;t)$ is the transport that has already happened in the 
past of the present time $t$, 
$a^{S}(s,t;t)$ is the transport to happen in the future of $t$,
and $a^{H}(s,t)$ is the net inter-gyre transport over all time.
The two properties of the TIME functions (Section~\ref{sec:chr})
are observed in this figure as follows. 
The temporal piece-wise independence property 
(\ref{eq:msum}) is observed by $a^{H}(s,t)=a^{U}(s,t; t)+a^{S}(s,t; t)$
for any $(s,t)$.
The invariance property (\ref{eq:adt}) is observed by 
$a^{H}(s,t)=a^{H}(s+\delta,t+\delta)$ for any $\delta$.

Due to the active $\mu^{H}(s,t)$ over the segment 
$[s_{{\rm J}},s_{{\rm S}}]$,
the displacement area functions have the relation
$a^{H}(s,t)\approx a^{S}(s,t; t)$
for $s<s_{{\rm J}}$.  
This is because no transport has happened there yet,
leading to $a^{U}(s,t; t)\approx 0$.
Similarly, $a^{H}(s,t)\approx a^{U}(s,t; t)$ holds
for $s>s_{{\rm S}}$  
because all transport has occurred already, leading to
$a^{S}(s,t; t)\approx 0$. 
Because the bi-infinite
displacement area function is the same as the Melnikov function,
an implication of these results to the Lagrangian transport is that
the development of the Lagrangian lobes 
are mainly governed by the flow dynamics between
$\overline{{\bf x}}^{H}(s_{{\rm J}})$ and
$\overline{{\bf x}}^{H}(s_{{\rm S}})$.
The Lagrangian lobes simply advect 
in the downstream direction of $\overline{{\bf x}}^{H}(s_{{\rm S}})$
without any further inter-gyre transport activities.

Figure~\ref{fg:dg_rl} shows the corresponding
displacement distance functions, 
$r^{U}(l,t; t)$, $r^{S}(l,t; t)$, and $r^{H}(l,t)$,
along the arc-length coordinate $l=l^{H}(s)$ measured in {\em km}.
The displacement distance functions are
inversely proportional to the local reference
velocity $|\overline{{\bf u}}(\overline{{\bf x}}^{H}(s))|$.
Therefore, the pseudo-lobes defined by the two consecutive zeros of the
displacement distance functions (see Appendix~\ref{sec:L})
are stretched and stay near zero around $s^{*}_{116}$
where $|\overline{{\bf u}}(\overline{{\bf x}}^{H}(s))|$ is maximum
along $\overline{W}^{H}$.
Accordingly it is near  $s=s^{*}_{116}$ that the inverse pseudo-PIP
$s^{\otimes}(t)$ associated with 
$r^{U}(l^{H}(s),t)$  and $r^{S}(l^{H}(s),t)$ 
exists for the pseudo-turnstile mechanism 
(see Appendix~\ref{sec:L}).
This also implies that the turnstile mechanism of the Lagrangian lobe
dynamics for the inter-gyre transport occurs near 
$\overline{{\bf x}}^{H}(s^{*}_{116})$.
In the downstream for large $s$,
all displacement distance functions show vertically elongated
pseudo-lobes due to 
very small $|\overline{{\bf u}}(\overline{{\bf x}}^{H}(s))|$.
The same phenomenon occurs in the upstream for small
$s$ near the upstream DHT.
This is related to the elongated structures of the Lagrangian lobes 
near the DHTs (Figure\ref{fg:dg_psi_mnf_C}b).
\fginsrt{fg:dg_rl}

The pseudo-lobes  of the bi-infinite displacement function can be used to
estimate the amount of the Lagrangian inter-gyre transport 
carried by the Lagrangian lobes
(Appendix~\ref{sec:L}).
Table~\ref{tbl:dg_EL} summarizes a quantitative comparison.
The area ${\cal A}({\cal L}^{L}_{j,j+1}(t))$ of 
the Lagrangian lobe ${\cal L}^{L}_{j,j+1}(t)$ 
is computed by the Lagrangian method
\cite{coulliette_wiggins_npg00},
while the area ${A}({L}^{H}_{j,j+1}(t)))$
of the pseudo-lobe ${L}^{H}_{j,j+1}(t)$  
is computed by (\ref{eq:AL}).
\tblinsrt{tbl:dg_EL}
Agreement between the two methods is quite good.

Clearly the Lagrangian lobe dynamics method provides precise geometry
associated with particle transport 
that the TIME method cannot
(compare Figure~\ref{fg:dg_psi_mnf_C}b with
 Figure~\ref{fg:dg_rl}).
The computation of  ${\cal W}^{U}(t)$ and ${\cal W}^{S}(t)$ using a
velocity data set given on a grid can be
computationally intensive.
As described in \cite{msw}, it requires sophisticated spatial and
temporal interpolation schemes.
In Figure~\ref{fg:dg_psi_mnf_C}b,
${\cal W}^{U}(t)$ and ${\cal W}^{S}(t)$ have about 
5000 and 2000 particles, respectively.
With this relatively large number of particles, 
it is also computationally challenging to preserve the area of 
each Lagrangian lobe precisely, due to the geometrical complexity 
of ${\cal W}^{U}(t)$ and ${\cal W}^{S}(t)$ as well as
the limited accuracy of the numerical schemes. 
In contrast, the TIME method is extremely efficient in computing the
transport because it requires a very small number of the simple operations.
In this example, we used only 1001 data points along $\overline{W}^{H}$
to compute $a^{H}(s,t)$. Once $a^{H}(s,t)$ is obtained,  it
automatically provides the value  $a^{H}(s+\tau-t,\tau)$ for any
 $(s+\tau-t,\tau)$.

\section{Summary and concluding remarks}
\label{sec:cncl}

We have developed a mathematical framework for the estimation of 
{\bf T}ransport {\bf I}nduced by the {\bf M}ean-{\bf E}ddy 
interaction (TIME) 
for flow properties and fluid particles with emphasis on
two-dimensional unsteady geophysical flows, without the assumption
of incompressibility. 
The TIME method estimates the amount of Lagrangian transport across
the kinematically-defined Eulerian boundary by the appropriately 
chosen streamline of the reference flow (Section~\ref{sec:bg}). 
The TIME method is a hybrid combination of Lagrangian 
and Eulerian methods and is based on the dynamical systems approach.
It enables on to analyze unique features of transport that neither the Lagrangian nor Eulerian 
methods can provide.

By considering two different aspects of transport, we obtain the
accumulation function for flow properties, as well as the
displacement distance and area functions for fluid particles
(Section~\ref{sec:fnc}). 
The dynamical systems approach leads to the useful characteristics
such as invariance, independence and coherency of the geometry
(Section~\ref{sec:chr}).
In the companion paper \cite{ide_wiggins_pd06b}, we develop a framework
for the analysis of the transport process in which 
these characteristics play a key role.

The notion of pseudo-lobes is developed to describe the geometry
associated with TIME.
When a heteroclinic  is used as the Eulerian boundary and the time interval
for transport to take place is set over a bi-infinite time
interval, the pseudo-lobes are geometrically closely
related to the Lagrangian lobes of the associated invariant manifolds.
The novel turnstile mechanisms for Lagrangian transport can 
be carried over in the TIME method by taking the mirror image of
the pseudo-lobes in the upstream region of the
heteroclinic connection (Appendix~\ref{sec:L}). 
An application to an oceanic problem and
a comparison with the Lagrangian lobe dynamics studied by
\cite{coulliette_wiggins_npg00} 
(Section~\ref{sec:application}). 

The TIME method is designed  to augment and supplement  Lagrangian and
Eulerian transport methods by providing
the unique capability to analyze the underlying transport 
processes, as it will be shown in \cite{ide_wiggins_pd06b}.
The method can be applied to more genercal cases than
the application presented in this paper; 
the method itself does not require
the time periodicity of the flow filed  
or a heteroclinic connection in the  reference state.
Various applications and extensions of the Eulerian
transport theory, including three-dimensionality
\cite{wiggins_ide_prep06}, brings a new point of view and
direction to transport studies in geophysical flows.

\section*{Acknowledgements}

This research is supported by ONR Grant No.~ N00014-09-1-0418, (KI)
and ONR Grant No.~N00014-01-1-0769 (SW). We thank Dr. Ana Mancho for providing us with
computational codes used in her studies of transport in the
wind-driven double gyre. We also thank Dr.~Michal Branicki for
providing us with his computational codes and related technical
help for the computation of the invariant manifolds.

\setcounter{section}{0}
\renewcommand{\thesection}{\Alph{section}}

\section{Mathematical Background on Perturbation of Trajectories}
\label{sec:pert_just}

In this appendix we given the necessary mathematical background that the TIME functions are valid approximations to the quantities that they measure to leading order in  the size of the fluctuation around the reference state. Essentially, the result that we need is that trajectories of the full, time dependent velocity field are ''close'', in the sense of the size of the fluctuation, to trajectories of the reference state velocity field on the time intervals of interest. The TIME functions are integrals of functions involving the pieces of the decomposed velocity field, flow properties, and geometrical features of the chosen Eulerian curve $C$

Two distinct situations need to be considered. One is where the velocity field is defined by an analytical formula, discussed in Section \ref{sec:P}. The other is where the velocity field is defined as a data set, discussed in Section \ref{sec:P_data}.

\subsection{A Velocity Field Defined by an Analytical Formula}
\label{sec:P}

In this appendix we collect together the results that we use on
the approximation of trajectories of the reference flow by
trajectories of the flow consisting of the reference flow and the
fluctuations about the reference flow over appropriate time
intervals. These results are stated here for completeness and they
can be found in a number of references. See, e.g.,
\cite{melnikov_tmms63,holmes_siamjam80,greenspan_holmes_book83,guckenheimer_holmes_book83,salam_etal_ieee83,salam_sastry_ieee85,salam_siamjam87}.
These results deal with the case of time-periodic fluctuations.
However, the arguments and proofs for aperiodically time-dependent
fluctuations are the same and are discussed in
\cite{malhotra_wiggins_jnls98,samwig}.

First, we recall notation established in Section \ref{sec:bg}.

The velocity field expressed analytically as the sum of a steady
reference state and an unsteady fluctuation is given by:

\begin{eqnarray} \label{eq:uunstdy_app}
  \frac{d}{dt}{\bf x}= {\bf u} ({\bf x},t)
= \overline{{\bf u}} ({\bf x}) + \varepsilon \hat{{\bf u}}({\bf
x},t),
\end{eqnarray}

\noindent (note that the following perturbation results are valid
for $\bf x$ either two or three dimensional). We denote a
trajectory of (\ref{eq:uunstdy_app}) by:

\begin{equation}
{\bf x}(t;\overline{{\bf x}}^{C}(s_{0}),t_{0}).
\label{eq:traj_pert}
\end{equation}

The reference velocity field is given by:

\begin{eqnarray} \label{eq:s-def_app}
    \frac{d}{ds} \overline{{\bf x}}^{C} (s)
  &=&\overline{{\bf u}} (\overline{{\bf x}}^{C} (s)),
\end{eqnarray}

\noindent
We denote a trajectory of the reference velocity field by:

\begin{equation}
 \overline{{\bf x}}^{C} (s_0 - t_0 + t).
 \label{eq:traj_ref}
 \end{equation}

Basic results in the theory of ordinary differential equations say
that if (\ref{eq:uunstdy_app}) is a  $C^r$ (i.e. $r$ times
continuously differentiable) function of $t$, $x$, and
$\varepsilon$, then (\ref{eq:traj_pert}) is a $C^r$ function of
$t$, $s$, and $\varepsilon$. Hence, we can Taylor expand in any of
those variables.

In particular,  (\ref{eq:traj_pert}) can be substituted into
(\ref{eq:uunstdy}) and differentiated with respect to
$\varepsilon$. In this way we obtain ordinary differential
equations for the coefficients (which are functions of time) for
the different powers of $\varepsilon$. Following this procedure,
we obtain:

\begin{equation}
 {\bf x}(t;\overline{{\bf x}}^{C}(s_{0}),t_{0})= \overline{{\bf x}}^{C} (s_0 - t_0 + t)
 + \varepsilon \hat{{\bf x}}(t;s_{0},t_{0}) + {\cal O}(\varepsilon^2),
 \label{eq:expan}
 \end{equation}

 \noindent
 where $ \hat{{\bf x}}(t;s_{0},t_{0})$ satisfies the {\em first variational equation}:

 \begin{equation}
 \frac{d}{dt}  \hat{{\bf x}} = D_{{\bf x}}
  \overline{{\bf u}} (\overline{{\bf x}}^{C} (s_0 - t_0 + t)) \hat{{\bf x}}
 + \hat{{\bf u}}  (\overline{{\bf x}}^{C} (s_0 - t_0 + t), t),
\label{eq:first_var}
\end{equation}

\noindent which is a linear, inhomogeneous, differential equation.
It is clear that with more work one could derive differential
equations whose solutions are the coefficients of the higher order
terms in the $\varepsilon$ expansion of of the trajectories.
However, the first order term in $\varepsilon$ will be sufficient
for our purposes.

In order for our results to be valid we will need estimates on the
time interval for which (\ref{eq:traj_pert}) and
(\ref{eq:traj_ref}) are ${\cal O}(\varepsilon)$ close (as measured
in an appropriate norm, say the maximum of the Euclidean distance
between the two trajectories over the time interval of interest).

We state our first result on closeness over finite time intervals.

 \begin{proposition}[Finite Time Approximation.]
 Suppose $\vert  {\bf x}(t_0;\overline{{\bf x}}^{C}(s_{0}),t_{0}) -\overline{{\bf x}}^{C} (s_0) \vert = {\cal O}(\varepsilon)$. Then
 $\vert  {\bf x}(t;\overline{{\bf x}}^{C}(s_{0}),t_{0}) -\overline{{\bf x}}^{C} (s_0 - t_0 +t) \vert
 = {\cal O}(\varepsilon)$ for $\vert t - t_0 \vert = {\cal O}(1)$.
 \label{prop:finite_time}
 \end{proposition}

 \noindent
 This result is elementary and well-known, and is a consequence of a simple application of Gronwall's inequality that can be found in the references given above, as
 well as many texts dealing with perturbation results.

In Section \ref{sec:bg} we discussed the situation where the
(steady) reference flow contained a hyperbolic stagnation point,
denoted $\overline{{\bf x}}^{{\rm dht}}$, with the hyperbolic
stagnation point having stable and unstable manifolds, $
\overline{W}^{U}\left(\overline{{\bf x}}^{{\rm dht}}\right)$ and
$\overline{W}^{S}\left(\overline{{\bf x}}^{{\rm dht}}\right)$,
respectively. We next state a result that describes how this
hyperbolic structure persists under perturbation of the reference
flow by the fluctuation.

Let $B_{\varepsilon_0}$ denote the ball of radius $\varepsilon_0$
centered at $\overline{{\bf x}}^{{\rm dht}}$.  Let $
\overline{W}^{U}_{\rm loc}\left(\overline{{\bf x}}^{{\rm
dht}}\right)$ denote the component of the intersection of $
\overline{W}^{U}\left(\overline{{\bf x}}^{{\rm dht}}\right)$ with
$B_{\varepsilon_0}$ that contains $\overline{{\bf x}}^{{\rm
dht}}$. Similarly for $\overline{W}^{S}_{\rm
loc}\left(\overline{{\bf x}}^{{\rm dht}}\right)$. Then we have the
following well-known result.

\begin{proposition}[Persistence of Hyperbolic Structures]
There exists $\varepsilon_0$ sufficiently small such that for all $0 < \varepsilon \le \varepsilon_0$ (\ref{eq:uunstdy_app}) has a hyperbolic (time dependent) trajectory  ${\bf x}^{\rm dht}_\varepsilon (t) = \overline{\bf x}^{\rm dht} + {\cal O}(\varepsilon)$. Moreover, ${\bf x}^{\rm dht}_\varepsilon (t)$ has local stable and unstable manifolds, denoted
$ \overline{W}^{S}_{\rm loc}\left(\overline{{\bf x}}^{{\rm dht}}_\varepsilon (t)\right)$ and
$\overline{W}^{U}_{\rm loc}\left(\overline{{\bf x}}^{{\rm dht}}_\varepsilon (t)\right)$, that are $C^r$ $\varepsilon$ close to
$\overline{W}^{S}_{\rm loc}\left(\overline{{\bf x}}^{{\rm dht}}\right)$ and
$\overline{W}^{U}_{\rm loc}\left(\overline{{\bf x}}^{{\rm dht}}\right)$, respectively.
\label{prop:persist}
\end{proposition}

\noindent
Note that it is an ''infinite time'' result.

Using an argument that combines Proposition \ref{prop:finite_time}
and Proposition \ref{prop:persist}, the following result can be
proven.

\begin{proposition}[Approximation on Semi-Infinite Time Intervals]
Suppose we choose initial conditions of trajectories in the stable
and unstable manifolds of the hyperbolic trajectory at $t=t_0$:

\begin{eqnarray}
{\bf x}^{S}(t;\overline{{\bf x}}^{C}(s_{0}),t_{0}) &  \in  &  \overline{W}^{S}\left(\overline{{\bf x}}^{{\rm dht}}_\varepsilon (t_0)\right), \\
 {\bf x}^{U}(t;\overline{{\bf x}}^{C}(s_{0}),t_{0}) & \in &  \overline{W}^{U}\left(\overline{{\bf x}}^{{\rm dht}}_\varepsilon (t_0)\right)
\end{eqnarray}

\noindent
then

\begin{eqnarray}
{\bf x}^{S}(t;\overline{{\bf x}}^{C}(s_{0}),t_{0})  = && \overline{{\bf x}}^{C} (s_0 - t_0 + t)+ \varepsilon \hat{{\bf x}}^{S}(t;s_{0},t_{0})  +   {\cal O}(\varepsilon^2)  \nonumber \\ 
&& \mbox{uniformly in t for} \quad t \in [t_0, \infty) \nonumber \\
&& \nonumber \\
{\bf x}^{U}(t;\overline{{\bf x}}^{C}(s_{0}),t_{0}) = && \overline{{\bf x}}^{C} (s_0 - t_0 + t)+ \varepsilon \hat{{\bf x}}^{U}(t;s_{0},t_{0}) + {\cal O}(\varepsilon^2)  \nonumber \\
&& \mbox{uniformly in t for} \quad t \in (-\infty, t_0] \nonumber \\
\end{eqnarray}

\noindent where $\hat{{\bf x}}^{S}(t;s_{0},t_{0})$ and $\hat{{\bf
x}}^{S}(t;s_{0},t_{0})$ satisfy

\begin{eqnarray}
 \frac{d}{dt}  \hat{{\bf x}}^{S} & = & D_{\bf x} \overline{{\bf u}} (\overline{{\bf x}}^{C} (s_0 - t_0 + t))\hat{{\bf x}}^{S}
 + \hat{{\bf u}}  (\overline{{\bf x}}^{C} (s_0 - t_0 + t), t) \quad \mbox{for} \quad t \in [t_0, \infty), \nonumber \\
  \frac{d}{dt}  \hat{{\bf x}}^{U} &  = &  D_{\bf x} \overline{{\bf u}} (\overline{{\bf x}}^{C} (s_0 - t_0 + t))\hat{{\bf x}}^{U}
  + \hat{{\bf u}}  (\overline{{\bf x}}^{C} (s_0 - t_0 + t), t) \quad \mbox{for} \quad t \in(-\infty, t_0], \nonumber \\
\end{eqnarray}

\end{proposition}

Now we give a brief derivation of the integral formula for the
displacement area discussed in Section
\ref{sec:fnc_ref_displacement}. Writing ${\bf x}'(t;s_{0},t_{0})
=\varepsilon\hat{{\bf x}}(t;s_{0},t_{0})$, $r'(t;l_{0},t_{0})=
\varepsilon \hat{r}(t;l_{0},t_{0})$ and $a'(t;l_{0},t_{0})=
\varepsilon\hat{a}(t;l_{0},t_{0})$, (\ref{eq:rao-a}) takes the
form:

   \begin{eqnarray}
    \hat{a}(t;l_{0},t_{0})&\equiv&
  \hat{r}(t;l_{0},t_{0})|\overline{{\bf u}}(\overline{{\bf x}}^{C}(s))|~=
    \overline{{\bf u}}(\overline{{\bf x}}^{C}(s_{0}-t_{0}+t))
  \wedge \hat{{\bf x}}(t;s_{0},t_{0})~, \nonumber \\
      \label{eq:rao-a_hat}
   \end{eqnarray}

Differentiating (\ref{eq:rao-a_hat})  with respect to $\tau$ gives

\begin{eqnarray}
\frac{d}{d\tau} \hat{a}(\tau;l_{0},t_{0}) & = & {\rm trace}\{
D_{\bf x}\overline{{\bf u}}(\overline{{\bf x}}^{C}(s-t+\tau)) \} \
\hat{a}(\tau;l_{0},t_{0}) + \mu^{C}(s-t+\tau,\tau)~, \nonumber
\\\label{eq:daodt}
\end{eqnarray}

that has a clear resemblance to (\ref{eq:xunstdy}). Unlike
(\ref{eq:xunstdy}) that is two-dimensional for ${\bf
x}'(t;s_{0},t_{0})$, (\ref{eq:daodt}) is a scalar, linear ordinary
differential equation and hence can be solved  analytically. Doing
so with initial condition $a'(t_{0};l_{0},t_{0})=0$ gives
(\ref{eq:ao-a}).

\subsection{Velocity Field Defined by a Data Set}
\label{sec:P_data}

When the velocity field ${\bf u}({\bf x},t)$ is given as a data set, we need
to consider more carefully the applicability of the collection of mathematical
results discussed  above since the data set does not contain an explicit $\varepsilon$. Practically, this means we must first decide
on a particular decomposition of  ${\bf u}({\bf x},t)$ into
$\overline{{\bf u}}({\bf x},t)$ and $\varepsilon\hat{{\bf u}}({\bf
x},t)$. Although the choice of $\overline{{\bf u}}({\bf x})$ is
not unique, a natural choice  is the time average (mean) of
${\bf u}({\bf x},t)$. The geometrical structure of the flow defined by $\overline{{\bf u}}({\bf x})$, we can
make a choice for $C=\{\overline{{\bf x}}^{C}(s)\}$. The residual
becomes the unsteady fluctuation, i.e.,

\begin{eqnarray}
{\bf u}'({\bf x},t)&=&{\bf u}({\bf x},t)-\overline{{\bf u}}({\bf x},t)~.
\end{eqnarray}

Once a decomposition of the data set is chosen we then need to examine the
smallness of ${\bf u}'({\bf x},t)$ with respect to $\overline{{\bf
u}}({\bf x})$. As in Section~\ref{sec:fnc} and
Appendix~\ref{sec:P}, we do so for the separate types of $C$:
finite time interval, and semi- or bi-infinite time interval.

The essential result required is to show that the unsteady perturbation is actually small for a
finite-time interval $C$. There are numerous ways of doing this. One way is to consider the quantity:

\begin{eqnarray}\label{eq:dltC_st}
\alpha^{C}({\bf x},t)&=& \frac{|{\bf u}'({\bf
x},t)|}{|\overline{{\bf u}}({{\bf x}})|}
\end{eqnarray}

\noindent
where $\vert \cdot \vert$ denotes a convenient norm.  This quantity can be
 estimated numerically in a neighborhood of $C$ for a time interval of interest.

For either a semi-infinite or a bi-infinite time interval,
$\alpha^{C}({\bf x},t)$ defined by (\ref{eq:dltC_st}) will not generally be small at the endpoints of $C$ where we have 
$\overline{{\bf x}}^{{\rm dht}}$ since
${|\overline{{\bf u}}(\overline{{\bf x}}^{{\rm dht}})|}=0$.  However, this does not affect the validity of the TIME functions since the integrand of each function contains 
$\overline{{\bf u}} (\overline{{\bf x}}^{C} (s_0 - t_0 + t))$ which vanishes exponentially fast as 
$\overline{{\bf x}}^{{\rm dht}}$ is approached along $C$. The argument here is the same as the proof of absolute convergence of the Melnikov integrals given in the references at the beginning of this appendix. Effectively, the smallness of (\ref{eq:dltC_st}) only needs to be established along $C$ outside of a neighborhood of any endpoints of $C$ that are hyperbolic stagnation points.

\section{Relation to Lagrangian Transport}
\label{sec:L}

This appendix describes the similarities and differences between 
the Lagrangian lobe dynamics that is closely associated with the pseudo-lobe
dynamics of the TIME method along $\overline{W}^{H}$. All TIME
functions used in this section have the superscript $\{\cdot\}^{H}$, 
reflecting the fact that the Eulerian boundary $C$ we are considering is  
$\overline{W}^{H}$.
We begin by presenting a brief description of Lagrangian transport
from the dynamical systems point of view; details of the theory
can be found in \cite{wiggins_book92,malhotra_wiggins_jnls98,msw,samwig}.

\subsection{Overview of Lagrangian transport}
\label{sec:L_overview}

\begin{figure}[htb!]
\begin{center}
\includegraphics[width=12.cm]{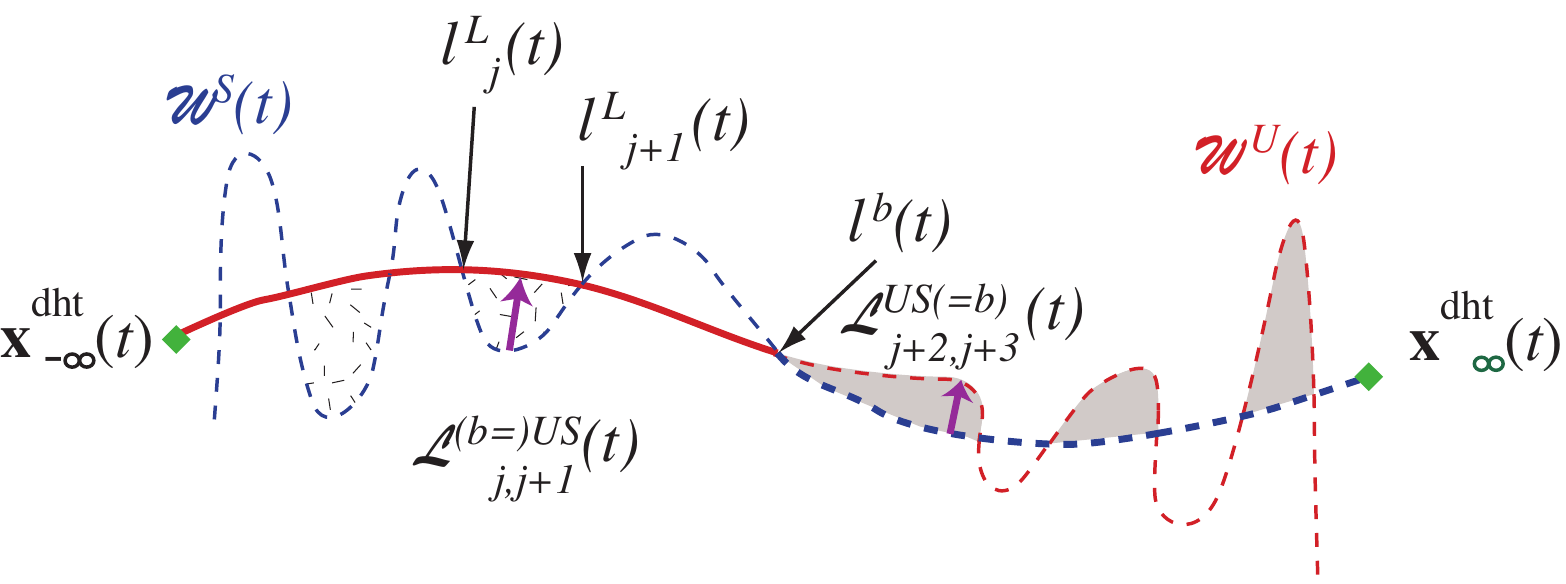}
\end{center}
\caption{Schematics of Lagrangian lobes in ${\bf x}$.
The shaded lobe ${\cal L}^{US(=b)}_{j+2,j+3}(t)$ corresponds to  fluid particles
which have already transported from the right to the left of the
deformable Lagrangian boundary ${\cal W}^{b}(t)$, 
while the swatched lobe ${\cal L}^{(b=)US}_{j,j+1}(t)$ corresponds to
fluid particles to be transported in the future time.}
\label{fg:Llobe}
\end{figure}

\textsf{Geometry of Lagrangian invariant manifolds.}
Lagrangian lobes are formed by time-dependent
unstable and stable invariant manifolds, ${\cal W}^{U}(t)$ 
and ${\cal W}^{S}(t)$,
of upstream and downstream time-dependent DHTs,
${\bf x}^{\rm dht}_{-\infty}(t)$ and
${\bf x}^{\rm dht}_{\infty}(t)$.
The geometry is schematically shown in  Figure~\ref{fg:Llobe}.
For convenience, we parameterize the coordinate of 
the points on ${\cal W}^{U}(t)$ by 
${\bf x}^{U}(s,t)$  and  on ${\cal W}^{S}(t)$ by ${\bf x}^{S}(s,t)$,
 respectively, by the flight-time coordinate $s$ using 
their normal projection onto $\overline{W}^{H}=\{\overline{{\bf x}}^{H}(s)\}$, 
i.e.,  
$({\bf x}^{U,S}(s,t)-\overline{{\bf x}}^{H} (s))\cdot
\mbox{\boldmath$\xi$\unboldmath}^{C}_{\parallel}(l^{C}(s))=0$
for both ${\cal W}^{U}(t)$ and ${\cal W}^{S}(t)$.
Then the normal distance from $\overline{W}^{H}$ to ${\cal W}^{U}(t)$
and  ${\cal W}^{S}(t)$  is:
 \begin{eqnarray}
  r_{\perp}^{U,S} (l^{H}(s),t) & = & 
 ({\bf x}^{U,S}(s,t)-\overline{{\bf x}}^{H} (s))
   \wedge  \mbox{\boldmath$\xi$\unboldmath}^{C}_{\parallel}(l^{C}(s))~,
\label{eq:rlUS-r}
 \end{eqnarray}
where the subscript $\{\cdot\}_{\perp}$ here represents the normal projection.
Accordingly, the normal distance from ${\cal W}^{S}(t)$ to ${\cal W}^{U}(t)$ is
defined by:
\begin{subequations}  \label{eq:rUS}
 \begin{eqnarray} \label{eq:rUS-exct}
  r_{\perp}^{L}(l,t) &=& r_{\perp}^{U}(l,t)-r_{\perp}^{S}(l,t).
 \end{eqnarray}
The superscript $\{\cdot\}^{L}$ stands for Lagrangian.
This normal distance $ r_{\perp}^{L}(l,t)$ is approximated 
using the so-called Melnikov function up to leading order,
which is identical to the displacement distance function of TIME:
 \begin{eqnarray} \label{eq:rUS-M}
  r^{L}(l^{H}(s),t)  &=&
 \frac{a^{H}(l^{H}(s),t)}{|\overline{{\bf u}}(\overline{{\bf x}}^{H}(s))|}
 = r^{H}(l^{H}(s),t).
 \end{eqnarray}
\end{subequations}
\fginsrt{fg:Llobe}

\textsf{Principal Intersection Point (PIP) sequence.}
An intersection sequence between ${\cal W}^{U}(t)$ and  ${\cal W}^{S}(t)$ 
can be expressed as a discrete sequence, $\{s^{L}_{j}(t)\}$ with
$s^{L}_{j}(t)<s^{L}_{j+1}(t)$.
These intersecting points are called principal intersection points
(PIPs).
The sequence can be transformed to the arc-length  coordinates
$\{l^{L}_{j}(t)\}$ using 
$l^{L}_{j}(t)= l^{H}(s^{L}_{j}(t))$.
Invariance of ${\cal W}^{U}(t)$ and ${\cal W}^{S}(t)$ guarantees  that
a trajectory starting at any PIP will remain a PIP for all time.
From (\ref{eq:rUS-M}), the PIP sequence can be approximated by
the zero sequence of $r^{H}(l^{H}(s),t)$ up to leading order:
\begin{eqnarray}\label{eq:lL-l}
s^{L}_{j}(t) \sim s^{H}_{j}(t), \qquad \qquad
l^{L}_{j}(t) \sim l^{H}(s^{H}_{j}(t)).
\end{eqnarray}

\textsf{Lagrangian lobe and its classification 
by the directionality of transport.}
A Lagrangian lobe, $\{{\cal L}^{L}_{j,j+1}(t)\}$, is then defined by
segments of ${\cal W}^{U}(t)$ and ${\cal W}^{S}(t)$ between a pair of
adjacent PIPs corresponding to $s^{L}_{j}(t)$ and
$s^{L}_{j+1}(t)$. We classify them into the two types, ${\cal L}^{US}$ or
${\cal L}^{SU}$, depending on whether the corresponding segment of
${\cal W}^{U}(t)$ lies to the left or to the right of the corresponding
segment of ${\cal W}^{S}(t)$, where the directionality is measured with
 respect to the forward direction of $s$ along $\overline{W}^{H}$.

\textsf{Area of a lobe.}
Using an elaborate computational scheme, it is possible to compute 
the area ${\cal A}({\cal L}^{L}_{j,j+1}(t))$  
 of ${\cal L}^{L}_{j,j+1}(t)$ \cite{coulliette_wiggins_npg00}.
Using (\ref{eq:rUS-M}) and (\ref{eq:lL-l}) as well as
 (\ref{eq:AL}) on $\overline{W}^{H}$, 
leading order approximation of ${\cal A}({\cal L}^{L}_{j,j+1}(t))$ 
is given by the Melnikov function  \cite{wiggins_book92}, 
which we denote by $A({\cal L}^{L}_{j,j+1}(t))$:
\begin{eqnarray}
{\cal A}({\cal L}^{L}_{j,j+1}(t))\approx
A({\cal L}^{L}_{j,j+1}(t))&=&A(L^{H}_{j,j+1}(t))~,
\end{eqnarray}
where $A(L^{H}_{j,j+1}(t))$ is given by (\ref{eq:AL}). 

\textsf{Moving boundary and boundary PIP.}
In order to describe Lagrangian transport, a Lagrangian boundary
must be defined using the segments of moving
invariant manifolds
${\cal W}^{U}(t)$  and ${\cal W}^{S}(t)$:
\begin{subequations} \label{eq:Wlb}
 \begin{eqnarray} \label{eq:Wrlb-W}
  {\cal W}^{b}(t) &=&
  \{{\bf x}^{U}(s,t), s\leq s^{b}(t)\}\cup \{{\bf x}^{S}(s,t), s\geq s^{b}(t)\},
 \end{eqnarray}
 where $s^{b}(t)\subset\{s^{L}_{j}(t)\}$ is called a boundary PIP
 as schematically shown in Figure~\ref{fg:Llobe}a.
Unlike Eulerian transport with a well-defined stationary boundary
$\overline{W}^{H}$,
the selection of $s^{b}(t)$ out of all the existing PIPs and hence the
selection of ${\cal W}^{b}(t)$ may not be unique.
However 
a physically meaningful
choice may be to have ${\cal W}^{b}(t)=\{{\bf x}^{b}(s,t)\}$
 geometrically 
''close''  to $\overline{W}^{H}$ 
\cite{romkedar_etal_jfm90}. This condition requires smallness of both 
$|r_{\perp}^{U} (l^{H}(s),t)|$ and $|r_{\perp}^{S} (l^{H}(s),t)|$ in the
neighborhood of $s^{b}(t)$, and hence smallness of 
$|r_{\perp}^{L} (l^{H}(s),t)|$ approximated by $|r^{L}(l^{H}(s),t)|$.
From (\ref{eq:rUS-M}), a reasonable choice of $s^{b}(t)$ is therefore
 \begin{eqnarray} \label{eq:Wrlb-l}
   s^{b}(t)&=&\{s\in s^{L}_{j}(t)~|~ min
 |l^{H}(s_{\overline{{\bf u}}})-l^{H}(s)|\}
 \end{eqnarray}
\end{subequations}
where
 $s_{\overline{{\bf u}}}
=\{s~|~max|\overline{{\bf u}}(\overline{{\bf  x}}^{H}(s))|\}$ 
is the maximum velocity point on $\overline{W}^{H}$.

\textsf{Further classification of ${\cal L}^{L}_{j,j+1}(t)$ 
by the timing of transport.} 
The relation between $s^{b}(t)$ and $s^{L}_{j}(t)$  leads to 
a further classification of the Lagrangian lobes 
beyond ${\cal L}^{US}$ or ${\cal L}^{SU}$ 
that also describes the timing of transport
associated with ${\cal L}^{L}_{j,j+1}(t)$.

Particles in ${\cal L}^{L}_{j,j+1}(t)$  with $s^{L}_{j}(t)<s^{b}(t)$  
are yet to cross ${\cal W}^{b}(t)$ and hence
have not involved in the transport process yet.
Because where ${\cal W}^{b}(t)$ is a segment of ${\cal W}^{U}(t)$,
we classify these Lagrangian lobes as ${\cal L}^{(b=)US}$ 
and ${\cal L}^{SU(=b)}$ with the same directionality as 
${\cal L}^{US}$ or ${\cal L}^{SU}$, respectively
On the contrary, 
particles in ${\cal L}^{L}_{j,j+1}(t)$  with $s^{L}_{j}(t)>s^{b}(t)$  
have crossed ${\cal W}^{b}(t)$ and hence
have already involved in the transport process.
We classify these Lagrangian lobes as  ${\cal L}^{US(=b)}$ and
 ${\cal L}^{(b=)SU}$ with the same directionality as
${\cal L}^{US}$ or ${\cal L}^{SU}$, respectively.

\textsf{Discrete transport time sequence.}
As particles advect downstream in the unsteady flow,
$s^{b}(t)$ must change in time not only by advection but also
for keeping the shape of ${\cal W}^{b}(t)$ as close as possible
to $\overline{W}^{H}$ as defined in (\ref{eq:Wrlb-l}).
Accordingly, $s^{b}(t)$ must jump from $s^{L}_{j}(t)$ to
$s^{L}_{j-\triangle j}(t)$ at some discrete time sequence,
 $\{t^{b}_{k}\}$
where $j^{b}_{k}$ is the corresponding PIP number and
 $\triangle j^{b}_{k}$ and increment for the $k$-th jump.
The integer  $\triangle j^{b}_{k}$ should be positive so  that the 
shift of $s^{b}(t)$ moves towards upstream at $t^{b}_{k}$.
Between two consecutive jumps $t\in(t^{b}_{k},t^{b}_{k+1})$,
there is no Lagrangian transport
between the kinematically-distinct regions
because they are defined by the same (moving) boundary
 ${\cal W}^{b}_{k}(t)$.
At  the discrete time sequence  $\{t^{b}_{k}\}$, Lagrangian transport
 by the lobe dynamics occur due to the change of
 the boundary.

\textsf{Turnstile mechanism of transport.}
At the discrete time sequence $\{t^{b}_{k}\}$,
a total  of $\triangle j^{b}_{k}-1$ lobes,
${\cal L}^{L}_{n,n+1}(t^{b}_{k})$, for
$n=(j^{b}_{k}-\triangle j^{b}_{k}), \ldots, (j^{b}_{k}-1)$,
go across ${\cal W}^{b}(t^{b}_{k})$ into the other region.
Therefore,
the lobes of ${\cal L}^{(b=)SU}$ turn into  ${\cal L}^{US(=b)}$
and those of ${\cal L}^{(b=)US}$ turn into  ${\cal L}^{SU(=b)}$.
This is the so-called turnstile mechanism.
As a consequence, Lagrangian transport
involves only a few lobes at a time that are spatially
confined near the time-dependent boundary point
$(s^{b}(t^{b}_{k}),t^{b}_{k})$.

\subsection{Pseudo-lobe dynamics and Lagrangian lobe dynamics}
\label{sec:L_geometry}

The turnstile mechanism of the Lagrangian method is crucial in
understanding the transport of particles between the
kinematically-distinct regions across the invariant boundary ${\cal W}^{b}(t)$.
Lagrangian transport occurs locally only near
$\overline{{\bf x}}^{H}(s^{b}(t^{b}_{k}),t^{b}_{k})$ as above.
at the discrete time sequence $\{t^{b}_{k}\}$. Therefore each
Lagrangian lobe ${\cal L}^{L}_{j,j+1}(t)$ can be classified decisively
whether transport associated with it has occurred or not, by
its relative position with respect to the $s^{b}(t)$. 

Table~\ref{tbl:compare-d} presents a general comparison between the TIME
and the Lagrangian method.
The TIME method computes the net amount of transport over
time across a stationary Eulerian boundary $C$.
At time $t$,
pseudo-lobes $\{L^{C}_{j,j+1}(t)\}$ may contain a mixture of transport
that occurred in the past  and in the future.
This effect particularly manifests itself 
along the heterolinic connection $\overline{W}^{H}$, 
for the past $(-\infty,t]$ and the future $\tau\in[t,\infty)$. 
This is because the transport boundary is fixed as $\overline{W}^{H}$ 
unlike ${\cal W}^{b}(t)$, 
which changes from ${\cal W}^{U}(t)$ to ${\cal W}^{S}(t)$ at $s^{b}(t)$. 
Therefore the TIME method by itself does not present such a mechanism
because it deals with the net amount of transport
over time but not individual particles.

\begin{figure}[!ht]
\begin{center}
\includegraphics[width=12.cm]{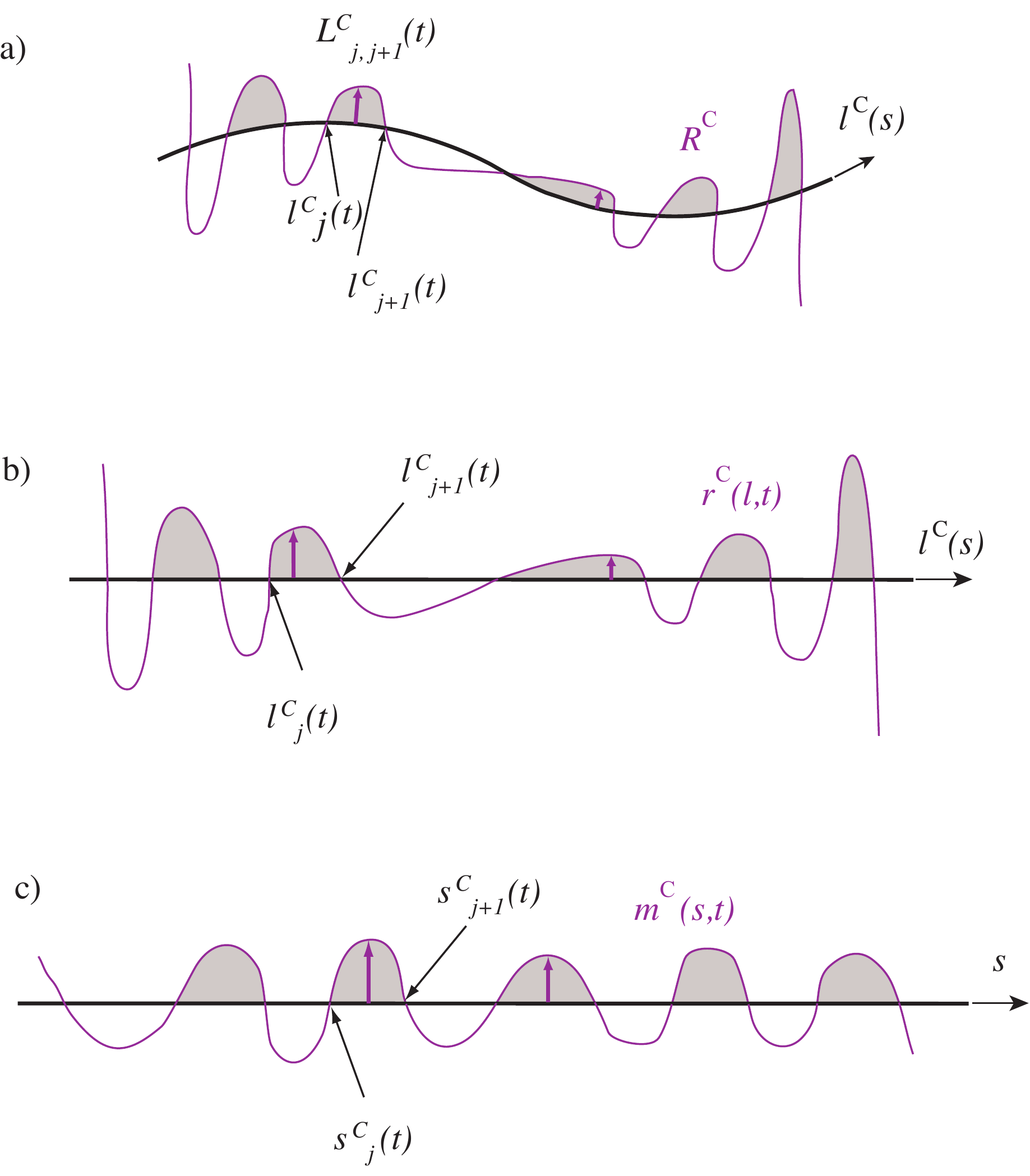}
\end{center}
\caption{Schematics of  pseudo-lobes along $C$: a) in ${\bf x}$; b) in $(l,r)$;
c) in $(s,m)$. The shaded areas correspond to the net accumulation
of fluid particles originally on the right side of $C$, where the
direction is defined with respect to the direction of increasing
$s$,
which accumulated
onto the left side of $C$ at time $t$. }
 \label{fg:Elobe}
\end{figure}

\begin{figure}[!ht]
\begin{center}
\includegraphics[width=12.cm]{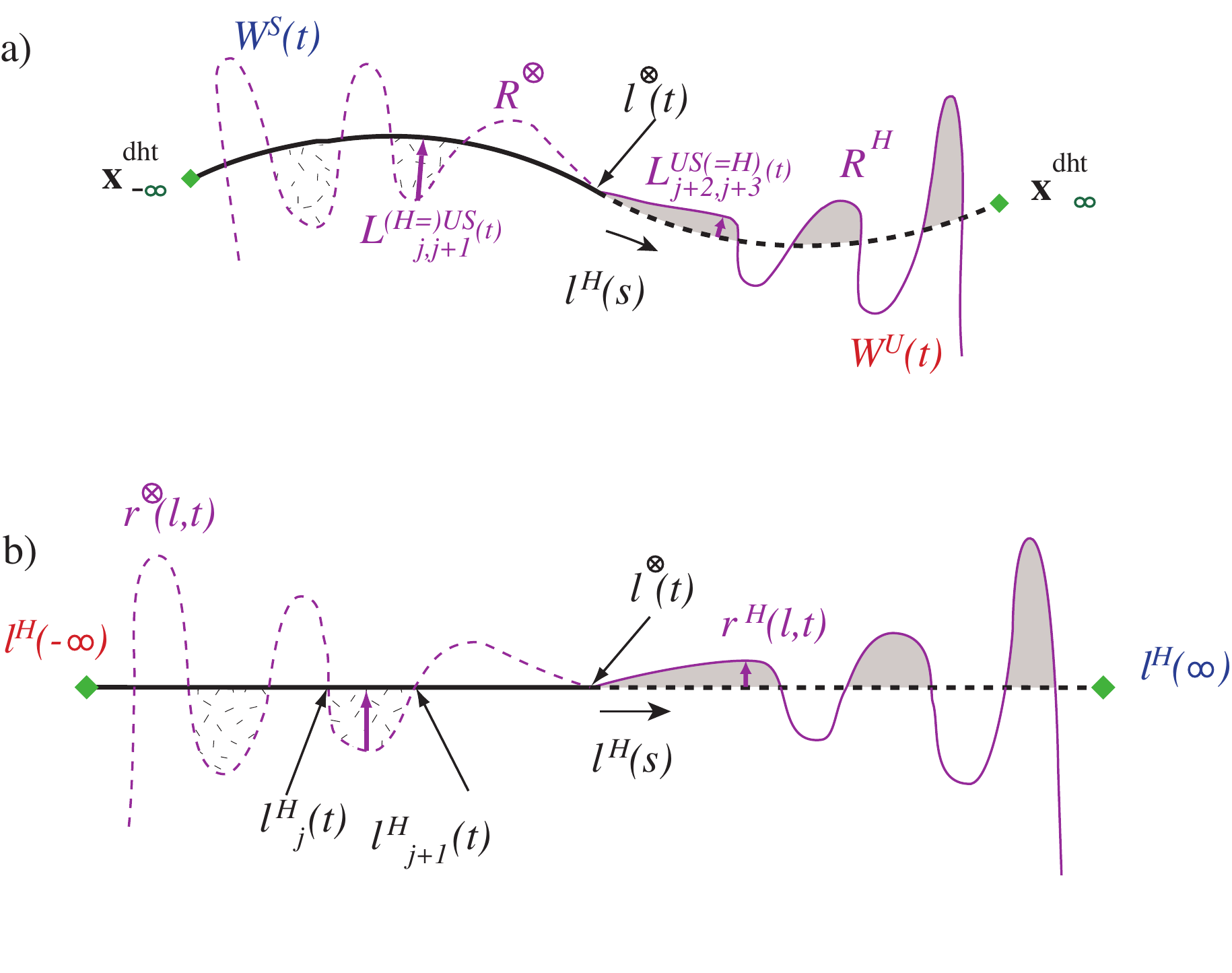}
\end{center}
\caption{Schematics of pseudo-Lagrangian lobe in a) ${\bf x}$ 
and b) $(l,r)$.
The shaded pseudo-Lagrangian lobe $L^{US(=H)}_{j+2,j+3}(t)$  
corresponds to net amount of fluid particles
which have already been transported from the right to the left of 
$\overline{W}^{H}$,
while the swatched lobe $L^{(H=)US}_{j,j+1}(t)$ corresponds to
the net amount of fluid particles
to be transported in the future time.}
\label{fg:plobe}
\end{figure}

A parallel development of the turnstile mechanism for the TIME method
is, however, possible by recognizing the nature of the TIME
method and realizing the meaning of the Lagrangian transport boundary
${\cal W}^{b}(t)$.
It requires the following modification to $R^{H}(t)$ as follows,
where $R^{H}(t)$ is defined  
in (\ref{eq:Rt}) along $\overline{W}^{H}$. In the downstream
direction where transport has happened and $\overline{W}^{H}$ corresponds
to ${\cal W}^{S}(t)$ associated with the downstream DHT, ${\cal W}^{U}(t)$
should correspond to $R^{H}(t)$. In the upstream direction where
transport is yet to happen and $\overline{W}^{H}$ corresponds to ${\cal W}^{U}(t)$
associated with the upstream DHT, however, ${\cal W}^{S}(t)$ should
correspond to the mirror image of $R^{H}(t)$:
\begin{eqnarray} \label{eq:RX}
R^{\otimes}(t) &=& \{(l,r) | r=-r^{H}(l,t)\}~
\end{eqnarray}
see Figure~\ref{fg:plobe}, in comparison with Figure~\ref{fg:Elobe}.
Table~\ref{tbl:compare_s} gives the summary of the definitions below.
\fginsrt{fg:plobe}
\tblinsrt{tbl:compare_s}

\textsf{Pseudo-manifolds.}
 We define the unstable pseudo-manifold $W^{U}(t)$ and stable
 pseudo-manifold $W^{S}(t)$ by the segments of $\overline{W}^{H}$,
 $R^{H}(t)$,  and  $R^{\otimes}(t)$ as follows. In the arc-length coordinate,
 $W^{U}(t)=\{(l,r)~|~(l,r^{U}(l,t))\}$ and
 $W^{S}(t)=\{(l,r)~|~(l,r^{S}(l,t))\}$ are defined by
\begin{subequations} \label{eq:cWRX}
\begin{eqnarray}
 r^{U}(l,t) & = & \left\{ \begin{array}{lll}
      0, & \mbox{for $l\leq l^{H}(s^{\otimes}(t))$} & \mbox{ on $\overline{W}^{H}$} \\
      r^{H}(l,t), & \mbox{for $l>^{H}(s^{\otimes}(t))$} &\mbox{ on $R^{H}(t)$}
                               \end{array} \right.
         \label{eq:cWRX-R}\\
 r^{S}(l,t) & = & \left\{ \begin{array}{lll}
      -r^{H}(l,t), & \mbox{for $l\leq l^{H}(s^{\otimes}(t))$}
         &\mbox{on $R^{\otimes}(t)$} \\
      0, & \mbox{for $l>l^{H}(s^{\otimes}(t))$} & \mbox{ on $\overline{W}^{H}$} \\
                               \end{array} \right. ~,
         \label{eq:cWRX-X}
\end{eqnarray}
where definition of $s^{\otimes}(t)$ is given below.
The distance $r^{U}(t)-r^{S}(t)$ between $W^{U}(t)$ and
 ${W}^{S}(t)$ is $r^{H}(t)$ by these definitions.
 It also  corresponds to $r^{L}(l^{H}(s),t)$ of the Lagrangian
 transport (\ref{eq:rUS-M}).
\end{subequations}

\textsf{Pseudo-boundary PIP.}
 We select $s^{\otimes}(t)$  to be
 \begin{eqnarray} \label{eq:sX-sB}
  s^{\otimes}(t) &\sim&s^{b}(t)~
 \end{eqnarray}
out of all the existing $\{s^{H}_{j}(t)\}$
 so that the resulting $W^{U}(t)$ and $W^{S}(t)$  are
 geometrically closest to their corresponding Lagrangian invariant
 manifolds ${\cal W}^{U}(t)$ and ${\cal W}^{S}(t)$.
This choice of $ s^{\otimes}(t)$ implies large
$|\overline{{\bf u}}(\overline{{\bf x}}^{H}(s^{b}(t)))|$
from (\ref{eq:Wrlb-l}) and hence small 
$|r^{H}(l^{H}(s^{\otimes}(t)),t)|$.

\textsf{Pseudo-Lagrangian lobes.}
Originally the pseudo-lobes are defined for $C$ in Section~\ref{sec:chr_geo}
 by the segments of $R^{C}(t)$ and $C$.
For transport dynamics across $\overline{W}^{H}$, however, we follow the
 convention of the Lagrangian lobes defined in Section~\ref{sec:L_overview}.
The pseudo-Lagrangian lobe 
is defined by segments of $W^{U}(t)$ and $W^{S}(t)$ between a pair of
adjacent pseudo-PIPs corresponding to $s^{H}_{j}(t)$ and $s^{H}_{j+1}(t)$.
We classify them into two types, $L^{US}$ or $L^{SU}$,
depending on whether the corresponding segment of $W^{U}(t)$ lies to
 the left or to the right of the corresponding segment of $W^{S}(t)$
 in the forward direction of $s$ along the boundary $\overline{W}^{H}$.
Table~\ref{tbl:dg_EL} was computed by following this convention.

\textsf{Directionality of transport across $\overline{W}^{H}$ by the
pseudo-Lagrangian lobes.} Like Lagrangian lobe dynamics,  the
relative position of
 $s^{H}_{j}(t)$  with respect to $s^{\otimes}(t)$ determines whether or not
the $j$-th pseudo-Lagrangian lobe 
has been transported across $\overline{W}^{H}$.
This leads to a further classification of pseudo-Lagrangian
lobes,
$L^{(H=)US}$ and $L^{US(=H)}$ for $L^{US}$  as well a
$L^{(H=)SU}$ and $L^{SU(=H)}$ for $L^{SU}$.
The pseudo-Lagrangian lobes of $L^{(H=)US}$ and $L^{SU(=H)}$  lie
 before $s^{\otimes}(t)$ and will cross $\overline{W}^{H}$ in the future
 $\tau>t$.
In contrast, the pseudo-Lagrangian lobes of $L^{US(=H)}$ and
 $L^{(H=)SU}$  lie after $s^{\otimes}(t)$  and   have crossed $\overline{W}^{H}$ in the past
 $\tau<t$.

\textsf{Discrete transport time sequence of TIME.}
 As in the case of the boundary PIP $s^{b}(t)$,
 the inverse pseudo-PIPs $s^{\otimes}(t)$
 also forms a discrete time sequence $\{t^{\otimes}_{k}\}$.
 By the construction of $s^{\otimes}(t)$ (\ref{eq:sX-sB}), 
$\{t^{\otimes}_{k}\}$ is an approximation of
 the discrete time sequence of  the boundary PIPs $\{t^{b}_{k}\}$, i.e.,
   $t^{\otimes}_{k} \sim t^{b}_{k}$
 up to leading order.

\textsf{Turnstile mechanism of the pseudo-Lagrangian lobes.}
Like the Lagrangian lobe dynamics,
at the discrete time sequence $\{t^{b}_{k}\}$,
a total  of $\triangle j^{\otimes}_{k}-1$ pseudo-Lagrangian lobes
go across $\overline{W}^{H}$ into the other region.
Therefore,
the lobes of $L^{(H=)SU}$ turn into  $L^{US(=H)}$
and those of $L^{(H=)US}$ turn into  $L^{SU(=H)}$.
This is the so-called pseudo-turnstile mechanism of the TIME method.

\newpage
\bibliographystyle{elsart-num}

\newpage
\addcontentsline{toc}{section}{\protect\numberline{}{List of Tables}}
\listoftables

\newpage
\addcontentsline{toc}{section}{\protect\numberline{}{List of Figures}}
\listoffigures

\clearpage
\addcontentsline{toc}{section}{\protect\numberline{}{Tables}}
\begin{table}[htb!]
\noindent
\begin{center}
\begin{tabular}{|c|ll|l|} \hline
Symbol & \multicolumn{2}{l|}{Definition} & Equations\\ \hline
$s$ & \multicolumn{2}{l|}{flight-time coordinate along $C$} 
 & (\ref{eq:s-def}) \\
$C=\{\overline{{\bf x}}^{C}(s)\}$ & 
 \multicolumn{2}{l|}{Eulerian boundary over the segment
 $s\in[s_{a},s_{b}]$, including:} 
 & (\ref{eq:s-def}) \\
 & \ $\quad \overline{W}^{S}$ & 
   semi-infinite for $s\in[s_{0},\infty)$ & (\ref{eq:WUS})\\
 & \ $\quad \overline{W}^{U}$ & 
   semi-infinite for $s\in(-\infty,s_{0}]$ & (\ref{eq:WUS})\\
 & \  $\quad \overline{W}^{H}$ & 
   bi-infinite for $s\in(-\infty,\infty)$ & (\ref{eq:WH})\\
$(s,t)=(s_{0}-t_{0}+t,t)$ & 
 \multicolumn{2}{l|}{reference particle trajectory (advection)
   along $C$}
 & (\ref{eq:xmn-sgm}) \\
$(l,r)=(l^{C}(s),r)$ & 
 \multicolumn{2}{l|}{arc-length coordinate in two dimensions} 
 & (\ref{eq:veca}) \\ \hline
$m^{C}(s,t;t_{0}:t_{1})$ & 
 \multicolumn{2}{l|}{accumulation function}
 & (\ref{eq:mint}),(\ref{eq:mrapst})-(\ref{eq:mrainf}) \\
$a^{C}(s,t;t_{0}:t_{1})$ & 
  \multicolumn{2}{l|}{displacement area function} 
  & (\ref{eq:rao}),(\ref{eq:mrapst})-(\ref{eq:mrainf}) \\
$r^{C}(s,t;t_{0}:t_{1})$ & 
 \multicolumn{2}{l|}{displacement distance function} 
  & (\ref{eq:rao}),(\ref{eq:mrapst})-(\ref{eq:mrainf}) \\ \hline
$s^{C}_{j}(t)$ &  \multicolumn{2}{l|}{pseudo-PIP}
  & (\ref{eq:sCjt}) \\
$L^{C}_{j,j+1}(t)$ &  \multicolumn{2}{l|}{pseudo-lobe, which belongs to
 the following two types:}
  & (\ref{eq:LCjjt}) \\
 &  $\qquad L^{RC}$ & from right to left across $C$ &  \\
 &  $\qquad L^{CR}$ & from left to right across $C$ &  \\ 
$A(L^{C}_{j,j+1}(t))$ &  \multicolumn{2}{l|}{signed area of pseudo-lobe}
  & (\ref{eq:AL}) \\ 
\hline
\end{tabular}
\caption{Glossary of TIME functions.  See text for details.}
\label{tbl:glssry}
\end{center}
\end{table}

\clearpage
\begin{table}[htb!]
\begin{center}
\begin{tabular}{|c||c|c|} \hline
      & \multicolumn{1}{|c|}{TIME}
      & \multicolumn{1}{c|}{Lagrangian} \\ \hline \hline
lobe no & $A(L^{H}_{j,j+1}(t_{B}))$ & ${\cal A}({\cal L}^{L}_{j,j+1}(t_{B}))$ \\
id. ($j$) &  (km$^{2}$)  & (km$^{2}$) \\ \hline
1 &  6858 &   6258  \\ \hline
2 & -6776 &  -6355  \\ \hline
3 &  6796 &   6201  \\ \hline
4 & -6783 &  -6216  \\ \hline
5 &  6760 &   6424 \\ \hline
6 & -6838 &  -6515  \\ \hline
7 &  6819 &   6696  \\ \hline
8 & -6811 &  -6119  \\ \hline
\end{tabular}
\caption{Quantitative comparison of transport by the TIME pseudo-Lagrangian
lobes using the pseudo-manifold described in Appendix~\ref{sec:L}
and  by the Lagrangian lobes using the method described in 
\cite{coulliette_wiggins_npg00} at $t^{*}_{944}$
(see Figure~\ref{fg:dg_psi_mnf_C}b for the lobe number).
 Results of the TIME pseudo-Lagrangian lobes are based on 
 the pseudo-manifolds described in Appendix~\ref{sec:L}.
 The signed area of a pseudo-lobe and a Lagrangian lobe corresponds to
 the amount of transport: a positive value represents transport from
 the subpolar gyre to the subtropical gyre, while a negative value
 represents transport from the subtropical gyre to the subpolar gyre.
}
\label{tbl:dg_EL}
\end{center}
\end{table}

\clearpage
\begin{table}
\begin{center}
\begin{tabular}{|c|c||c|c|} \hline
 \multicolumn{2}{|c||}{} & TIME & Lagrangian \\ \hline \hline
 \multicolumn{2}{|c||}{type} 
  & particle, flow property $Q$ & particle \\ \hline
boundary & spatial & any reasonable $C$
     & ${\cal W}^{b}(t)$ only \\ \cline{2-4}
           &  temporal & stationary & continuously deformable \\
          &  & &
    \& discontinuous at $\{t^{b}_{k}\}$\\ \hline
${\bf u}({\bf x},t)$ & restriction &
   $|{\bf u}'({\bf x},t)|$ small along $C$ & no restriction \\ \cline{2-4}
    & spatial requirement 
     & along $C$ only & non-local in ${\bf x}$ \\ \cline{2-4}
   & {temporal period}   & flexible & infinite \\ \hline
 transport  & temporal & continues & turnstile at $\{t^{b}_{k}\}$ \\ \cline{2-4}
 process & spatial & all along $C$ 
  & locally near  ${\bf x}^{b}(l^{b}(t^{b}_{k}),t^{b}_{k})$  \\ \hline
 \multicolumn{2}{|c||}{computation} & efficient & exhaustive \\ \hline 
 \multicolumn{2}{|c||}{accuracy}    & leading order & accurate \\ \hline
\end{tabular}
\caption{A general comparison between the TIME method and Lagrangian method.}
\label{tbl:compare-d}
\end{center}
\end{table}

\clearpage
\begin{table}[htb!]
\begin{center}
\begin{tabular}{|c|c|c|c|c|} \hline
  \multicolumn{3}{|c|}{ }  & TIME & Lagrangian \\ \hline
 \multicolumn{2}{|c|}{ }& {intersection sequence} 
  &$\{s^{H}_{j}(t)\}$  
 &$\{s^{L}_{j}(t)\}$   \\ 
  \multicolumn{2}{|c|}{geometry} & {distance}  & $r^{H}(l,t)$ & $r^{L}(l,t)$ \\
 \multicolumn{2}{|c|}{ } &{area }     & $A(L^{H}_{j,j+1}(t))$
   & ${\cal A}({\cal L}^{L}_{j,j+1}(t))\approx A({\cal L}^{L}_{j,j+1}(t))$ \\ \hline
 \multicolumn{2}{|c|}{}& {boundary point} & $s^{\otimes}(t)$ & $s^{b}(t)$ \\
  \multicolumn{2}{|c|}{transport by} &{time sequence} 
        & $\{t^{\otimes}_{k}\}$ & $\{t^{b}_{k}\}$ \\
  \multicolumn{2}{|c|}{turnstile} & {lobe number} 
     & $j^{\otimes}_{k}$ & $j^{b}_{k}$ \\
 \multicolumn{2}{|c|}{ } & {lobe increment} & $\triangle j^{\otimes}_{k}$
        & $\triangle  j^{b}_{k}$ \\ \hline
(pseudo-) & right & future& $L^{(H=)US}$ & ${\cal L}^{(b=)US}$  \\
lobe & to left & past &  $L^{US(=H)}$  & ${\cal L}^{US(=b)}$  \\ \cline{2-5}
type & left  & future & $L^{SU(=H)}$ & ${\cal L}^{SU(=b)}$ \\
 & to right & past & $L^{(H=)SU}$ & ${\cal L}^{(b=)SU}$ \\ \hline
\end{tabular}
\caption{Relation between the TIME method
 and Lagrangian lobes for transport associated with $\overline{W}^{H}$.
 For (pseudo-)lobe type, ``right to left'' and ``left to right''
 indicate the direction of transport while ``future'' and ``past''
 indicate the timing of transport with respect to the present time $t$.
See text for details.}
\label{tbl:compare_s}
\end{center}
\end{table}

\clearpage
\addcontentsline{toc}{section}{\protect\numberline{}{Figures}}

\newpage

\newpage

\newpage

\begin{figure}[!ht]
\begin{center}
\includegraphics[width=15.cm]{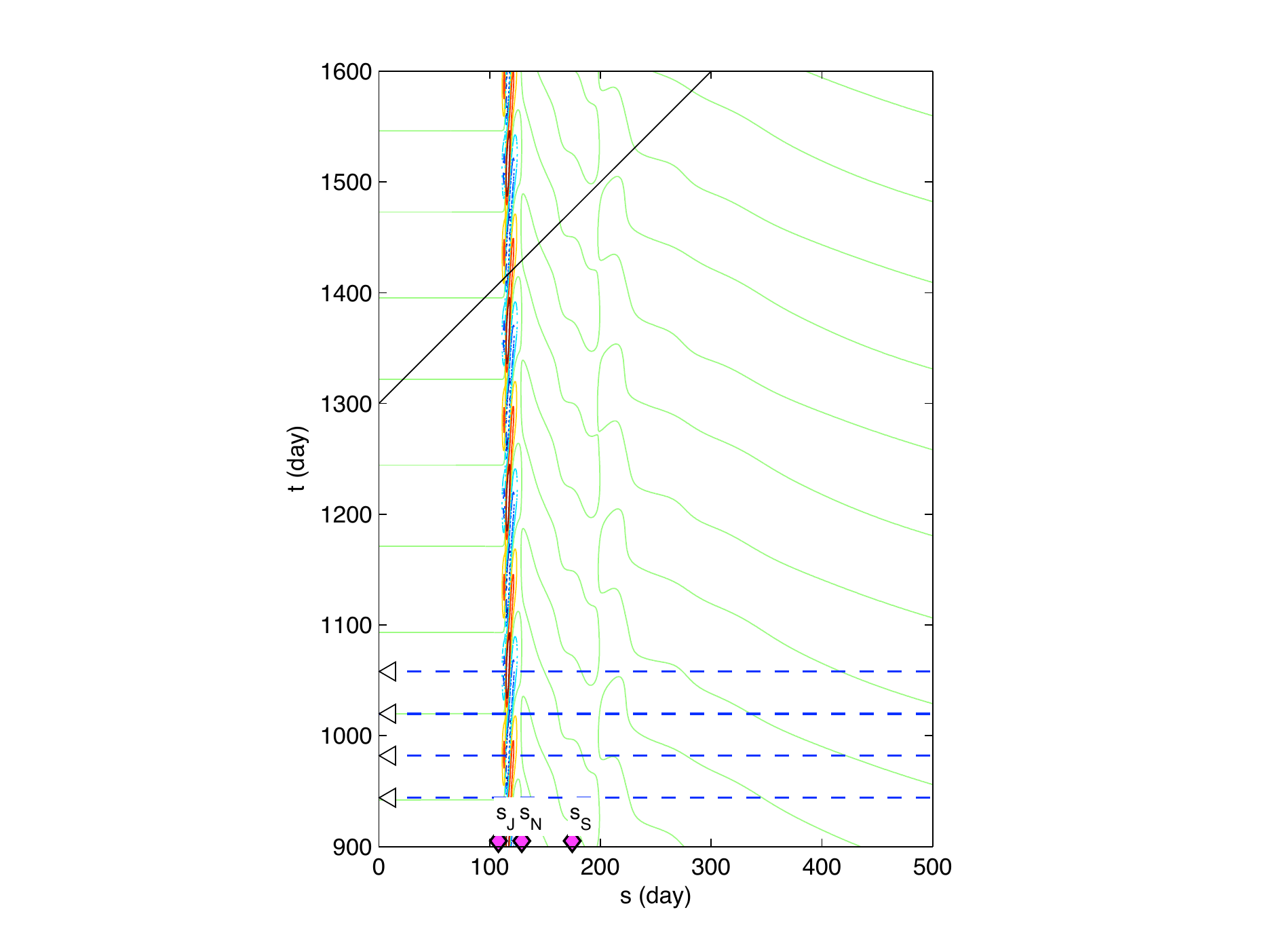}
\end{center}
\caption{The instantaneous flux function
$\mu^{H}(s,t)$ in the Hovm\"{o}ller diagram 
with dash-dot line for the negative values
and contour interval 40km$^{2}$/day.
The diagonal line is an example of reference trajectory going through
 $(s_{0},t_{0})=(0,1300)$. The horizontal lines corresponds to the four
 phases during one period of the oceanic oscillation,
at $t=$ $t^{*}_{944}$, $t^{*}_{944}+T/4$,
$t^{*}_{944}+T/2$, $t^{*}_{944}+3T/4$. }
\label{fg:dg_must}
\end{figure}

\newpage

\begin{figure}[!ht]
\begin{center}
\includegraphics[width=15.cm]{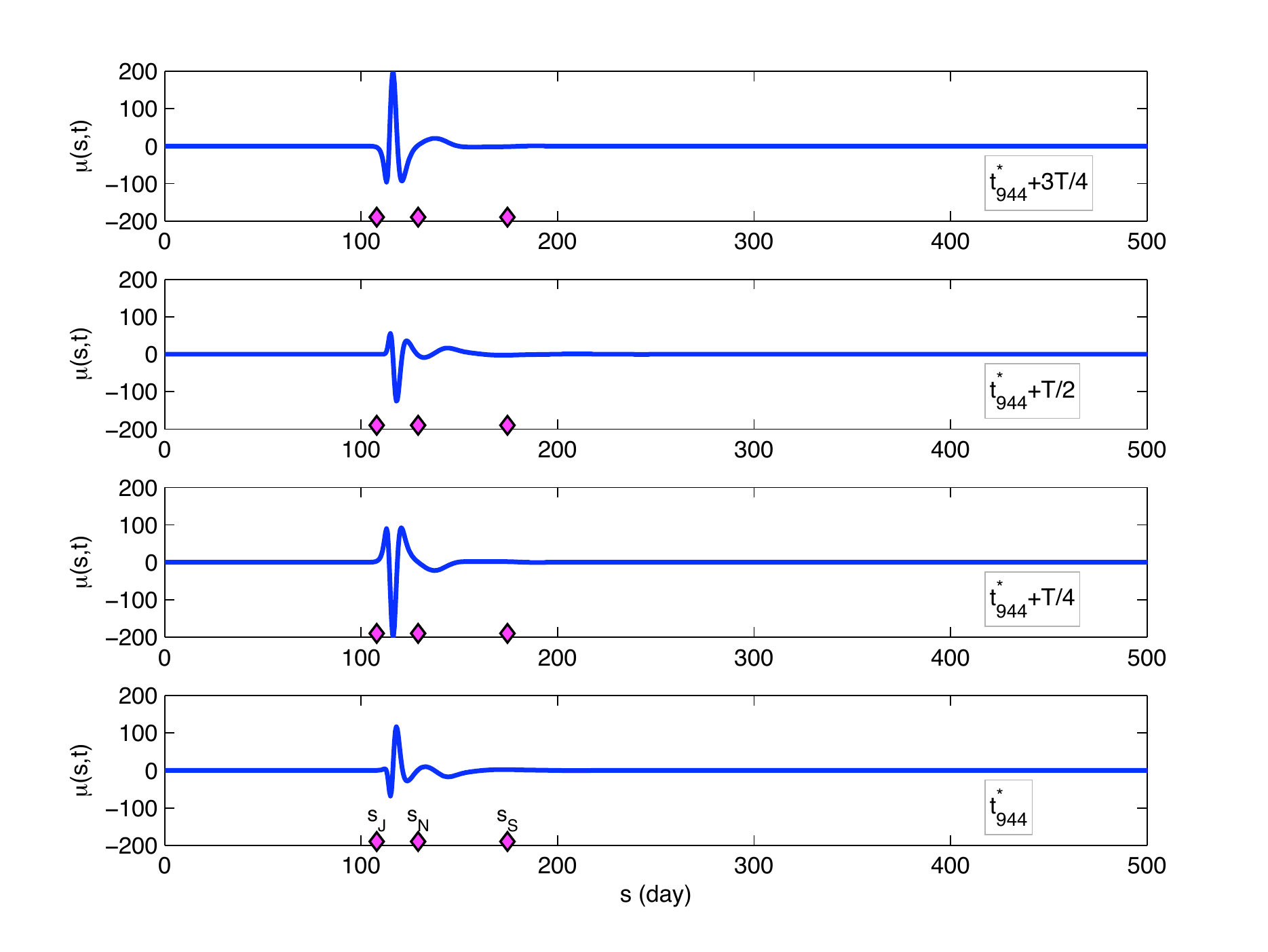}
\end{center}
\caption{The instantaneous flux function
$\mu^{H}(s,t)$ at $t=$ $t^{*}_{944}$, $t^{*}_{944}+T/4$,
$t^{*}_{944}+T/2$, $t^{*}_{944}+3T/4$  from bottom to top (right)
 with corresponding time indicated in Figure~\ref{fg:dg_must}.}
\label{fg:dg_mus}
\end{figure}

\newpage

\begin{figure}[!ht]
\begin{center}
\includegraphics[width=15.cm]{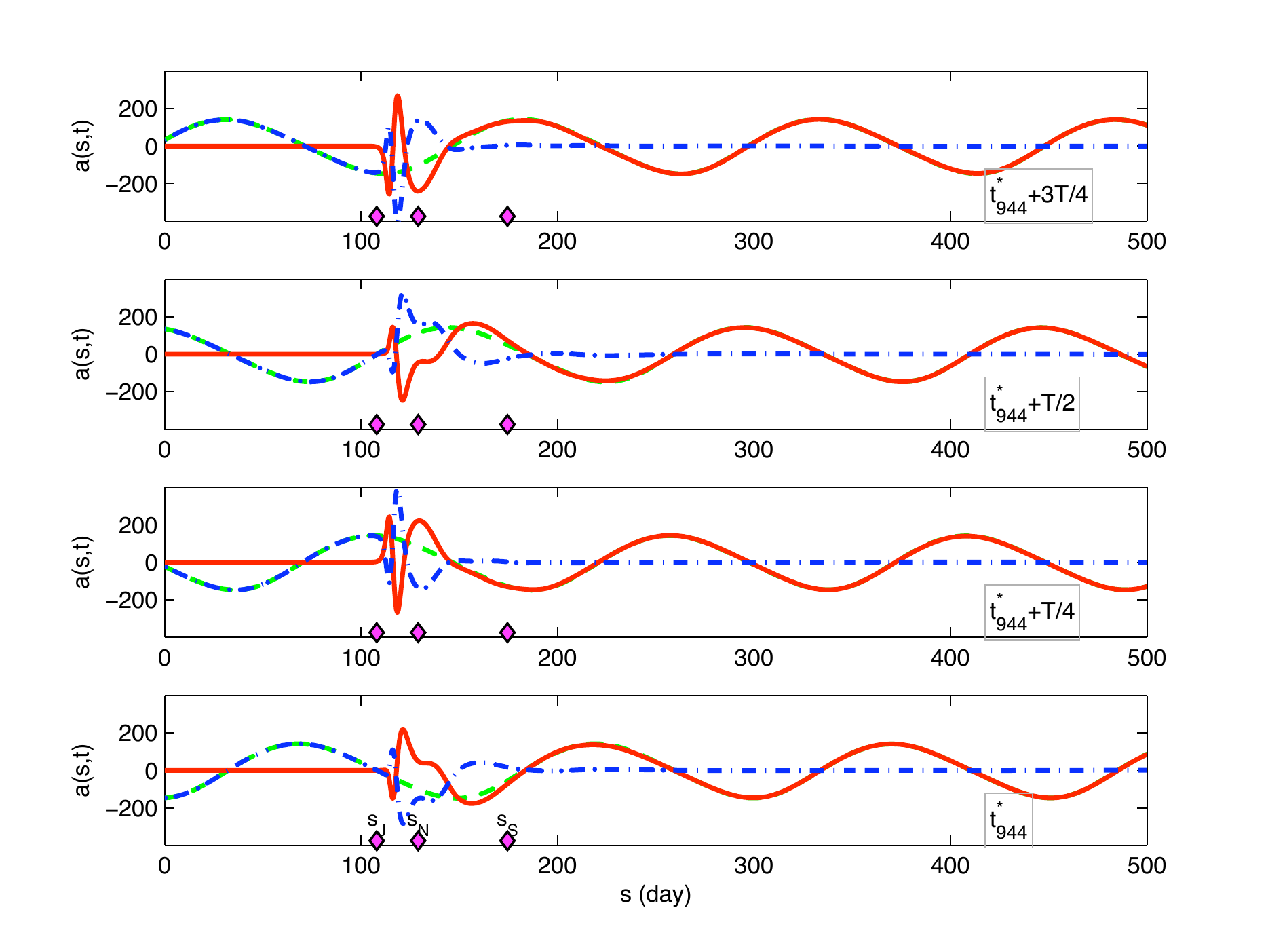}
\end{center}
\caption{The accumulation functions
$a^{H}(s,t)$ (dash line), $a^{U}(s,t;t)$ (solid line), 
and $a^{S}(s,t;t)$ (dash-dot line)  
at
$t^{*}_{944}$, $t^{*}_{982}(=t^{*}_{944}+T/4)$, 
$t^{*}_{1020}(=t^{*}_{944}+T/2)$,  
and $t^{*}_{1058}(=t^{*}_{944}+3T/4)$ (from bottom to top)
 with corresponding time indicated in Figure~\ref{fg:dg_must}:
$a^{H}(s,t)$ almost overlaps $a^{S}(s,t;t)$ for $s<s_{{\rm J}}$
and $a^{U}(s,t;t)$ for $s>s_{{\rm S}}$.
}
\label{fg:dg_ms}
\end{figure}

\newpage

\begin{figure}[!ht]
\begin{center}
\includegraphics[width=15.cm]{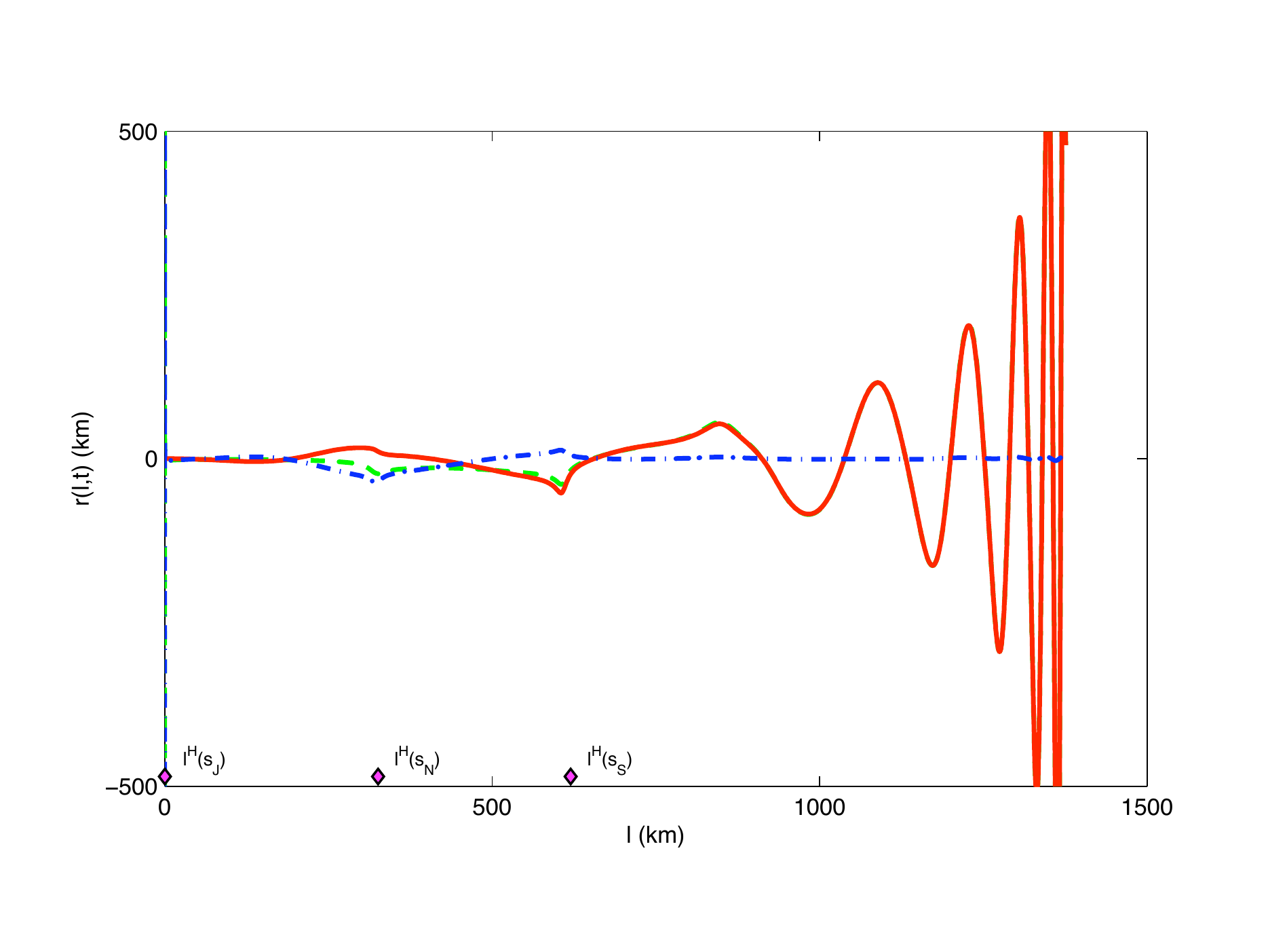}
\end{center}
\caption{The displacement distance functions, 
$r^{H}(l,t)$ (dash line), $r^{U}(l,t;t)$ (solid line), 
and $r^{S}(l,t;t)$ (dash-dot line)
at $t^{*}_{944}$ corresponding to the bottom panel in
Figure~\ref{fg:dg_ms};
$r^{H}(l,t)$ almost overlaps $r^{S}(l,t;t)$ for $l<l^{H}(s_{{\rm J}})$
and $r^{U}(s,t;t)$ for $l<l^{H}(s_{{\rm S}})$, although
pseudo-lobes of $r^{H}(l,t)$ and $r^{U}(l,t;t)$ accumulate near $l=0$
 and cannot be seen in this figure.}
\label{fg:dg_rl}
\end{figure}

\newpage

\newpage

\end{document}